\documentclass[journal=jacsat,manuscript=article]{achemso}
\usepackage{textcomp,gensymb,amssymb,amsmath,xcolor,physics}
\usepackage{pdfpages}
\author{Krishna Murali}
\affiliation[Equal]
{Equal contribution}
\alsoaffiliation[IIScECE]
{Department of Electrical Communication Engineering, Indian Institute of Science, Bangalore 560012, India}
\author{Medha Dandu}
\affiliation[Equal]
{Equal contribution}
\alsoaffiliation[IIScECE]
{Department of Electrical Communication Engineering, Indian Institute of Science, Bangalore 560012, India}
\author{Kenji Watanabe}
\affiliation[NIMS1]
{Research Center for Functional Materials, National Institute for Materials Science,
1-1 Namiki, Tsukuba 305-044, Japan}
\author{Takashi Taniguchi}
\affiliation[NIMS2]
{International Center for Materials Nanoarchitectonics, National Institute for Materials Science,
1-1 Namiki, Tsukuba, 305-044 Japan}
\author{Kausik Majumdar}
\email{Kausikm@iisc.ac.in}
\affiliation[IIScECE]
{Department of Electrical Communication Engineering, Indian Institute of Science, Bangalore 560012, India}

\title{Accurate Extraction of Schottky Barrier Height and Universality of Fermi Level De-pinning of van der Waals Contacts}

\begin{document}
\begin{abstract}
Due to Fermi level pinning (FLP), metal-semiconductor contact interfaces result in a Schottky barrier height (SBH), which is usually difficult to tune. This makes it challenging to efficiently inject both electrons and holes using the same metal - an essential requirement for several applications, including light-emitting devices and complementary logic. Interestingly, modulating the SBH in the Schottky-Mott limit of de-pinned van der Waals (vdW) contacts becomes possible. However, accurate extraction of the SBH is essential to exploit such contacts to their full potential. In this work, we propose a simple technique to accurately estimate the SBH at the vdW contact interfaces by circumventing several ambiguities associated with SBH extraction. Using this technique on several vdW contacts, including metallic 2H-TaSe$_2$, semi-metallic graphene, and degenerately doped semiconducting SnSe$_2$, we demonstrate that vdW contacts exhibit a universal de-pinned nature. Superior ambipolar carrier injection properties of vdW contacts are demonstrated (with Au contact as a reference) in two applications, namely, (a) pulsed electroluminescence from monolayer WS$_2$ using few-layer graphene (FLG) contact, and (b) efficient carrier injection to WS$_2$ and WSe$_2$ channels in both nFET and pFET modes using 2H-TaSe$_2$ contact.
\end{abstract}
\begin{figure}[!hbt]
\centering
%\vs{-0.1in}
%\hs{-1in}
\includegraphics[scale=0.5]{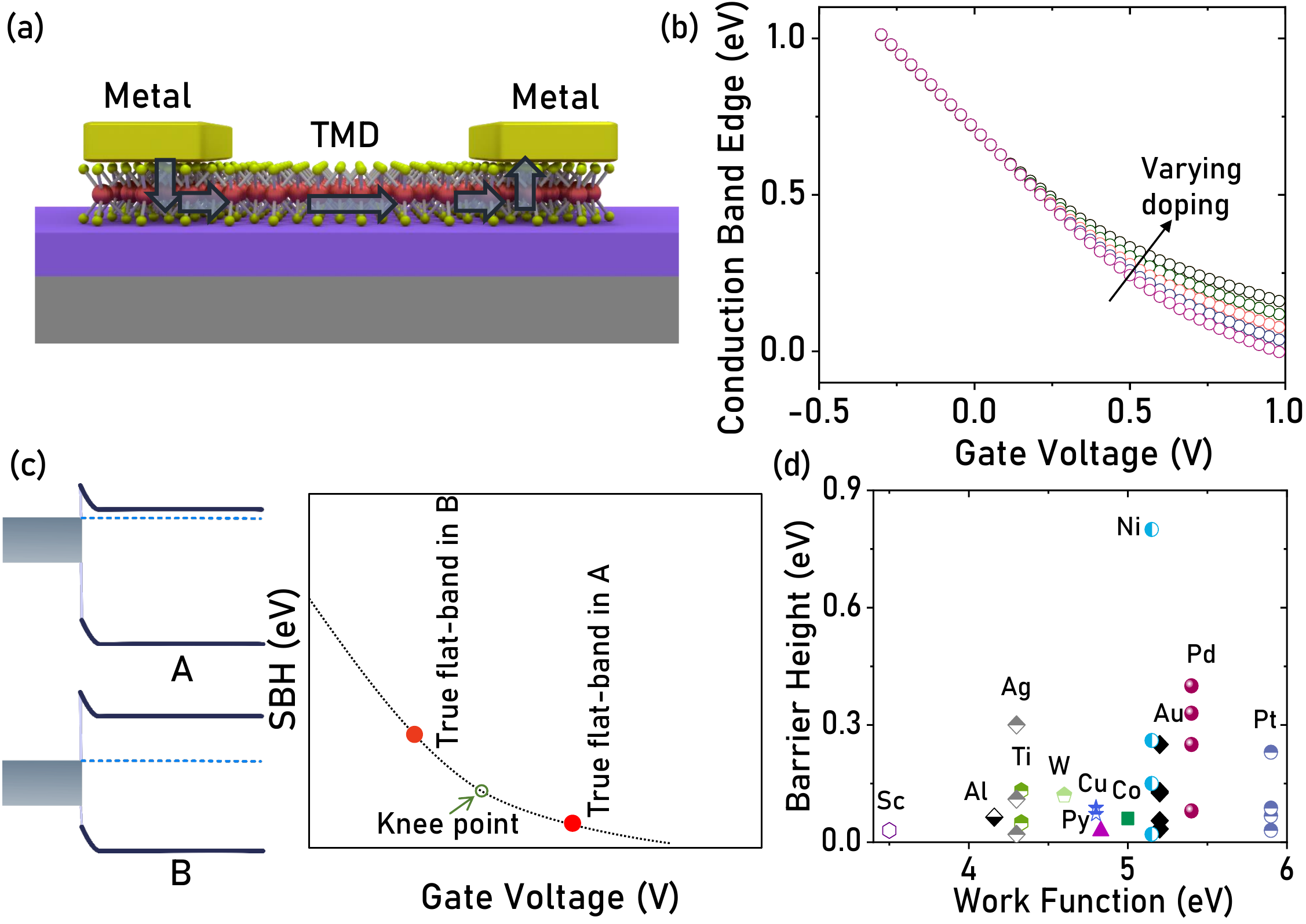}
%\vspace{-1.8in}
\caption{\textbf{Sources of ambiguities in SBH extraction:} (a) Schematic diagram of a back-gated lateral structure used for SBH extraction. The arrows indicate mixed dimensional carrier transport in the contact region. (b) The position of the conduction band edge (with respect to the Fermi level) as a function of V$_g$ for different channel doping, as obtained from the solution of coupled Poisson-Schrodinger equations. The channel thickness is taken as 10 nm. (c) A representative graph to illustrate the deviation of actual flat-band condition from the knee point (Green open circle) for cases of small SBH (A) and large SBH (B). (d) A benchmarking plot illustrating reported SBH values of different metals on MoS$_2$, indicating a large variation.}\label{Figures:Fig1}
\end{figure}
\begin{figure}[!hbt]
\centering
%\vs{-0.1in}
%\hs{-1in}
\includegraphics[scale=0.5]{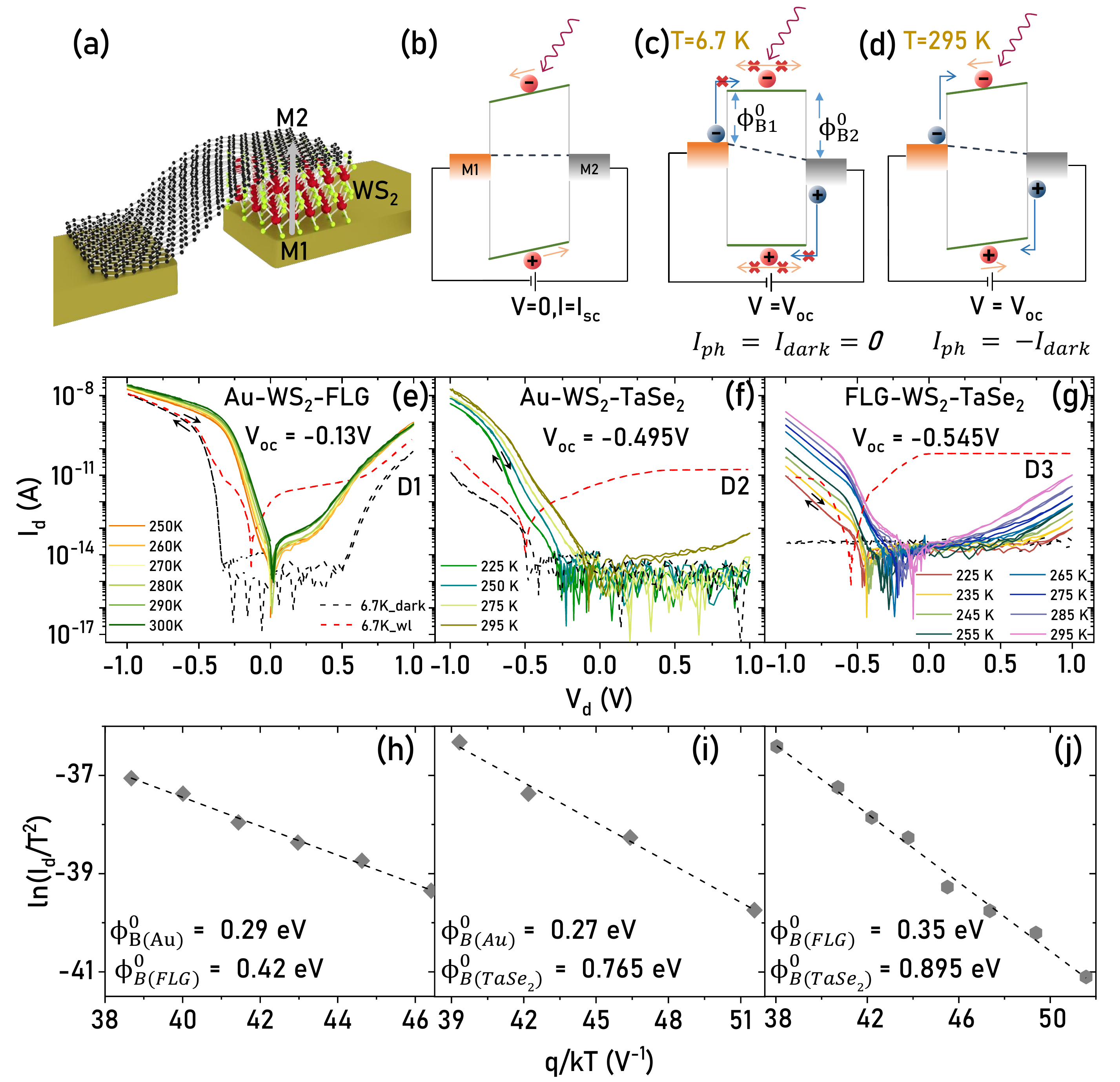}
%\vspace{-1.8in}
\caption{\textbf{Proposed SBH extraction method:} (a) Schematic of the proposed vertical test structure with asymmetric contacts. (b)-(d) Schematic representation of band diagrams (b) in equilibrium showing built-in electric field in the multi-layer, (c) ideal flat-band condition at $V_{oc}$ (at low temperature where $I_{dark}$ is suppressed), and (d) deviation from flat-band condition at $V_{oc}$ at higher temperatures due to non-zero $I_{dark}$. (e)-(g) Current-Voltage characteristics under dark condition at different temperatures for the three stacks, namely, Au-WS$_2$-FLG (D1), Au-WS$_2$-TaSe$_2$ (D2), and FLG-WS$_2$-TaSe$_2$ (D3). The black and the red dashed traces indicate current with and without light, respectively, at $T=6.7$ K. The $V_{oc}$ value at $6.7$ K in each structure is indicated in the inset, which indicates the true flat-band condition due to suppressed $I_{dark}$ at this temperature. (h)-(j) The corresponding Arrhenius plots for D1-D3 along with linear fits (dashed lines) to deduce the SBH of one of the contacts. The SBH of the other contact is obtained by adding the $V_{oc}(T=6.7$ K$)$.
 }\label{Figures:Fig2}
\end{figure}
\begin{figure}[!hbt]
\centering
%\vs{-0.1in}
%\hs{-1in}
\includegraphics[scale=0.4]{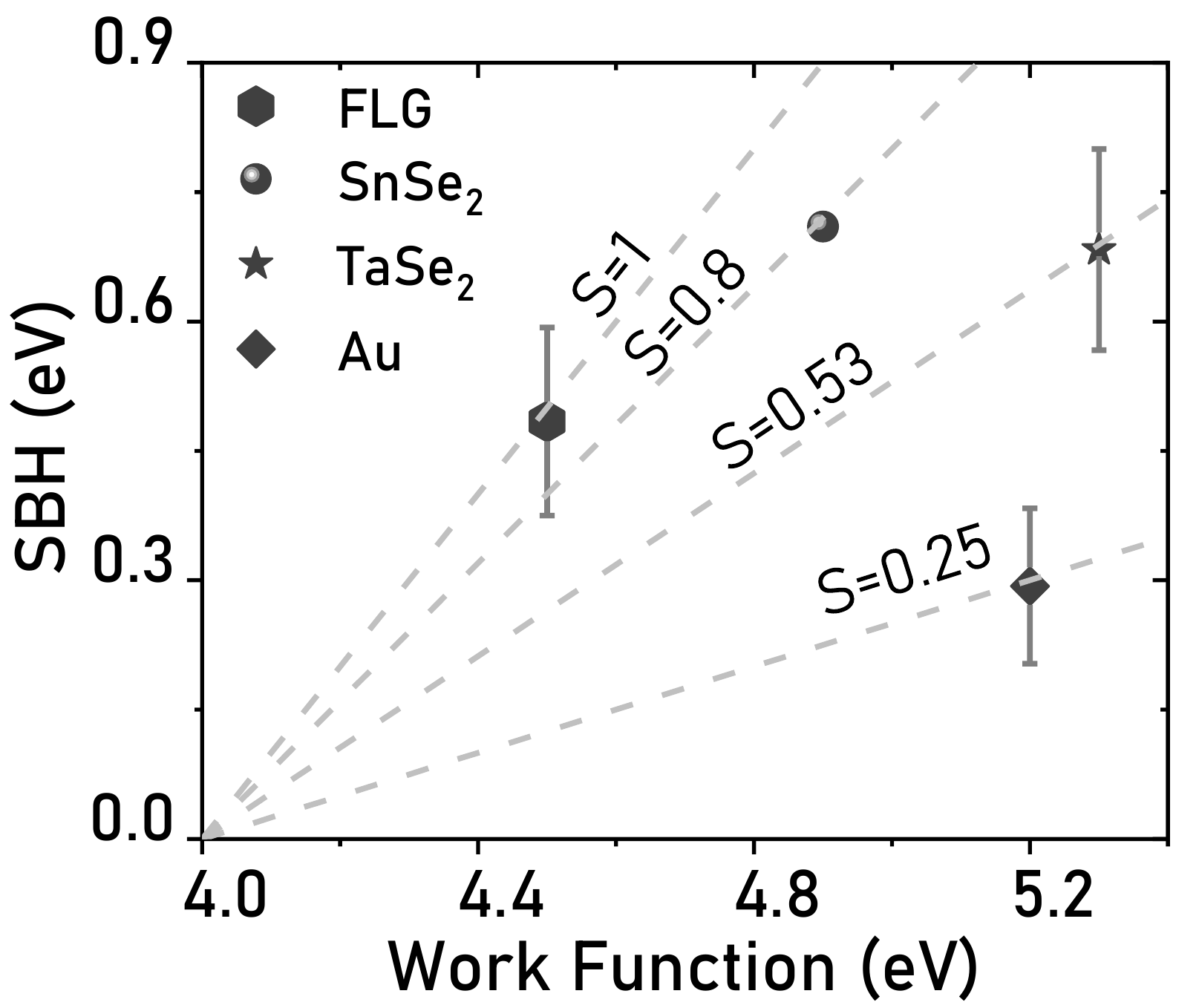}
%\vspace{-1.8in}
\caption{\textbf{Fermi level de-pinning with vdW contacts:} Summary of the extracted SBH versus metal work function Au, FLG, TaSe$_2$, and SnSe$_2$, when contacted to WS$_2$. The dashed lines correspond to the different pinning factors $(S)$. The vdW contacts exhibit significantly de-pinned nature compared to Au.}\label{Figures:Fig3}
\end{figure}
\begin{figure}[!hbt]
\centering
%\vs{-0.1in}
%\hs{-1in}
\includegraphics[scale=0.35]{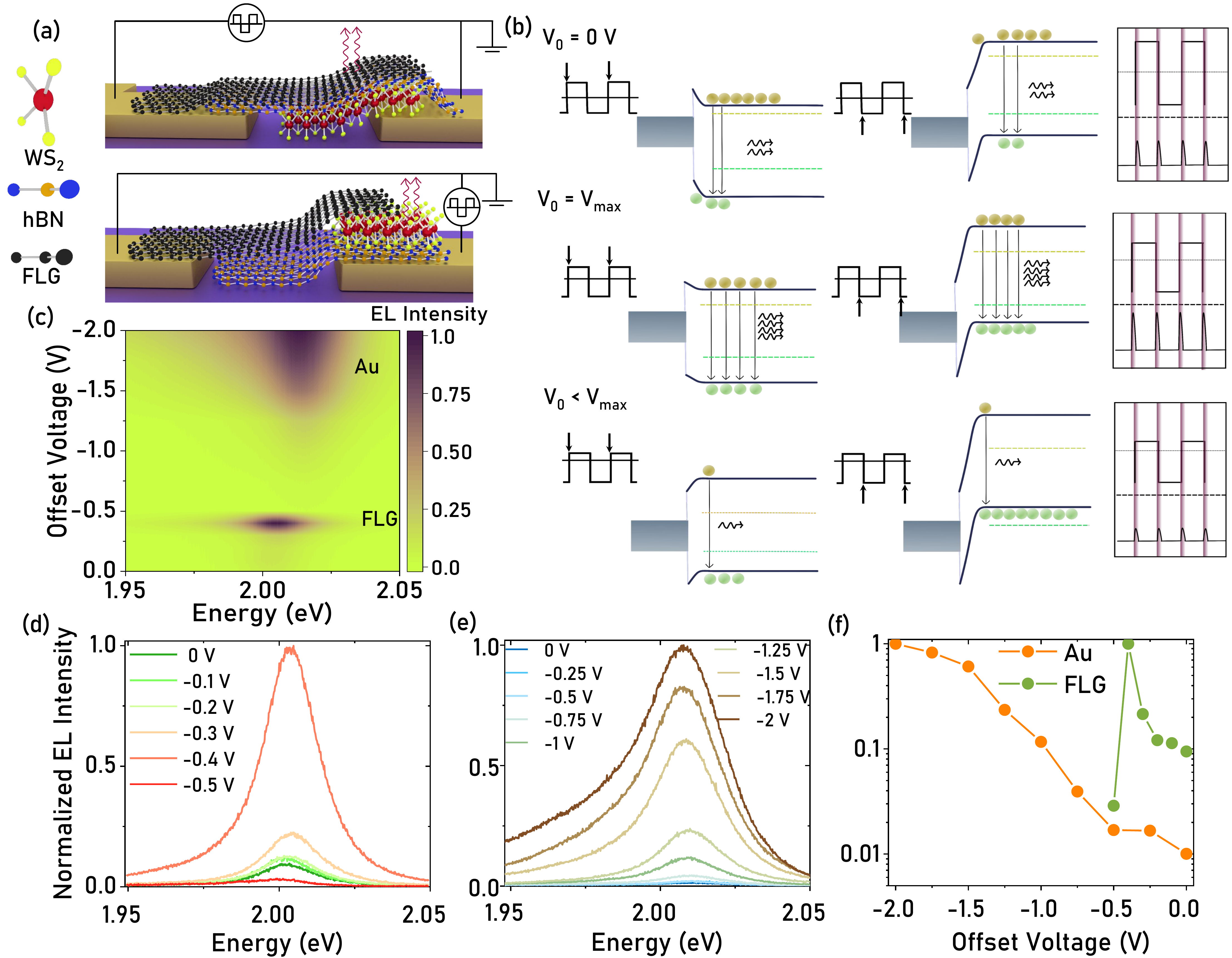}
%\vspace{-1.8in}
\caption{\textbf{Pulsed Electroluminescence - unipolar versus ambipolar injection:} (a) Schematic diagram of the pulsed electroluminescence devices with hBN as gate dielectric. Top panel shows the device with Au as source contact and bottom panel shows the device with FLG as source contacts. A pulse train ($5$ V peak-to-peak) riding on a dc offset voltage ($V_0$) is applied between gate and source electrode, and light emission occurs near the source contact edge. (b) Principle of operation and the role of $V_0$ in EL intensity modulation. The top, middle, and bottom rows correspond to $V_0=0$, $V_0=V_{max}$ (maximum EL intensity), and $V_0 < V_{max}$ for the case where Fermi level of the source is pinned close to the conduction band edge. (d) The colour plot of the measured EL intensity from 1L-WS$_2$ as a function of emission energy and $V_0$ for Au and FLG as the source contacts. $V_{max}$ occurs for FLG at a much lower voltage than for Au, indicating relative values of SBH for these two contacts. (e)-(f) Individual EL spectra of Figure (d) for different $V_0$ for (e) Au and (f) FLG contacted device. (f) Normalized peak EL intensity plotted in log scale as a function of $V_0$ for the two devices.}\label{Figures:Fig4}
\end{figure}
\begin{figure}[!hbt]
\centering
%\vs{-0.in}
\hspace{-0.5in}
\includegraphics[scale=0.325]{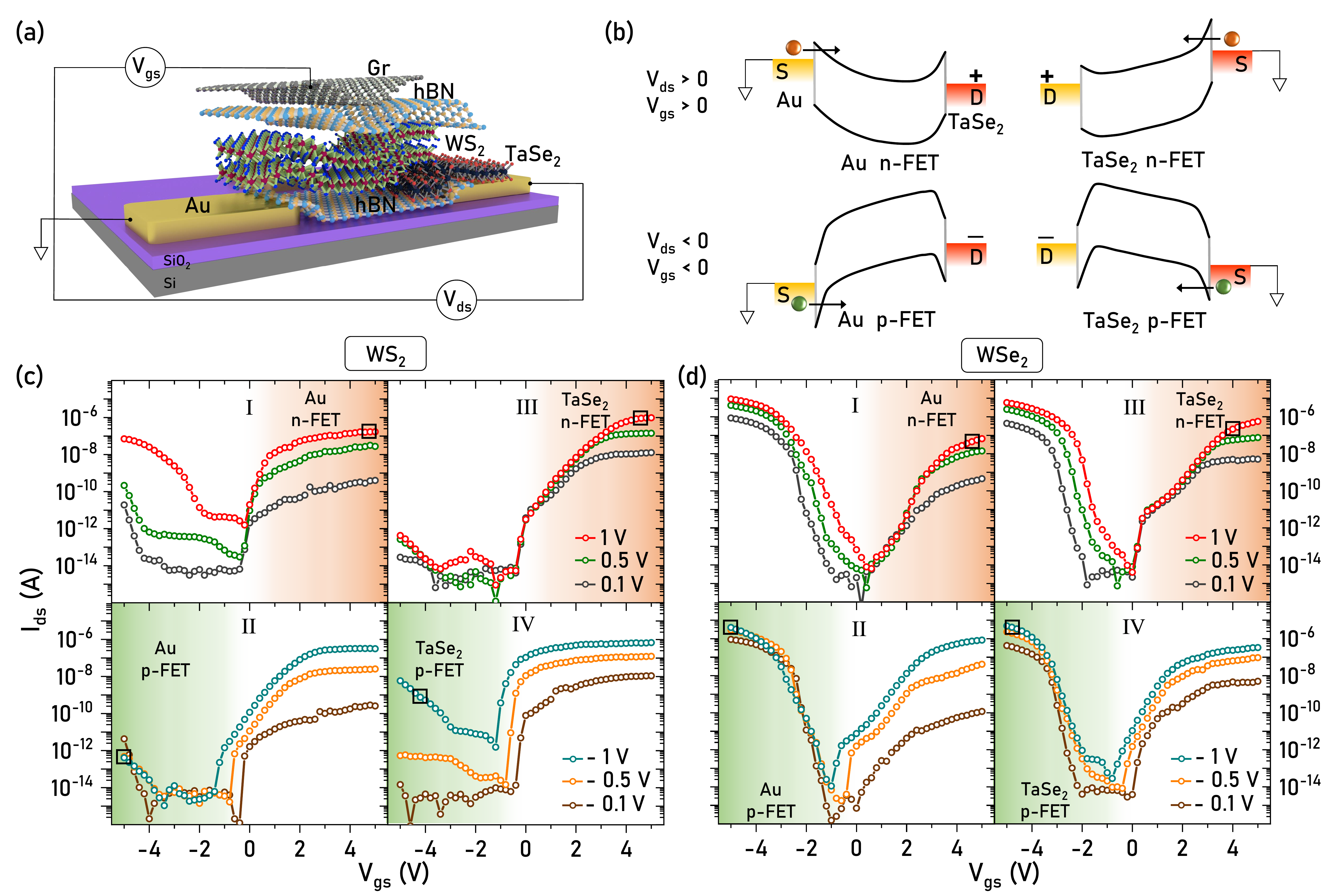}
%\hspace{-0.5in}
\caption{\textbf{Ambipolar injection from TaSe$_2$:} (a) Schematic of top-gated lateral FET with asymmetric contacts, Au and TaSe$_2$. The biasing configuration shown here is for TaSe$_2$ electron injection. (b) Band diagrams of carrier injection into the channel with Au and TaSe$_2$ as source in the MOSFET mode ($V_{gs}>0$ and $V_{ds}>0$ for n-FET; $V_{gs}<0$ and $V_{ds}<0$ for p-FET). (c) Transfer characteristics of top gated few layer WS$_2$ FET in log scale. Colored regions in orange and green highlight the electron and hole injection characteristics of Au (TaSe$_2$) source in panels I and II (III and IV), respectively. Black squares indicate the current levels at $|V_{ds}|= 1$ V for a gate overdrive of $5$ V and $3$ V in n-FET and p-FET modes, respectively. (d) Transfer characteristics of top gated few layer WSe$_2$ FET in log scale. Black squares highlight the current levels at $|V_{ds}|= 1$ V for gate overdrive of $4$ V.} \label{Figures:Fig5}
\end{figure}

\newpage
To be able to minimize the fractional contribution of the metal-semiconductor contact resistance in the total device resistance is a key to achieve high-performance electronic devices \cite{Ng1987,Giubileo2017,Taur2007,Schulman2018}. Unfortunately, with device scaling, this fractional contribution increases due to a reduction in the area of the contact footprint. The Schottky barrier height (SBH) at the metal-semiconductor interface plays a key role in determining the contact resistance. In an ideal contact interface, the SBH ($\phi_B^0$) is given by the difference between the metal work function ($W_m$) and the semiconductor electron affinity ($\chi_{s}$), as predicted by the Schottky-Mott rule \cite{Szeb}:
\begin{equation}\label{eq:SMRule}
\phi_B^0 = W_m - \chi_{s}
\end{equation}
However, in a real scenario, the interfacial dipole, metal induced gap states (MIGS), and interfacial defects lead to Fermi level pinning (FLP) \cite{Allain2015,Kim2017,Wang2020}, which has plagued the development of ultra-low resistance contacts for several materials, including n-Ge \cite{Gallacher2012,Nishimura2007}, p-InGaAs \cite{Lin2013,Yu1997}, and p-MoS$_2$ \cite{Zhang2019,Zhang2013,Chuang2014,Nipane2016,Liu2016}.
\\
In the context of the layered semiconductors, the possibility of ultra-clean, atomically smooth, dangling-bond-free surface, coupled with a relatively weak van der Waals (vdW) interaction at the interface reignited the hope to achieve a completely de-pinned contact close to the Schottky-Mott limit \cite{Liu2018,Song2020}. However, the estimated SBH value is often confounded by several ambiguities related to the extraction procedure. Given the scientific and technological importance, it is highly desirable to extract the ``true'' SBH values of the vdW contacts in an unambiguous manner. In this work, we propose a simple, yet powerful technique to achieve the same. We show that while the Au Fermi level is strongly pinned close to the conduction band edge of multi-layer WS$_2$, several types of vdW contacts including few-layer graphene (FLG), metallic 2H-TaSe$_2$ \cite{Mehak2019b}, and degenerately doped semiconducting SnSe$_2$ \cite{Krishna2018} exhibit highly de-pinned nature.
\\
Is a de-pinned contact always technologically desirable? Not necessarily in every situation. For example, in unipolar device applications, a contact strongly pinned close to the band edge is desirable to reduce SBH and hence the contact resistance, as in the case of n-InAs \cite{Singisetti2008,Shiraishi1994}, p-Ge \cite{Li2018,Spann2005}, and n-MoS$_2$ \cite{Das2013b,Chen2013}. Also, if the de-pinning is achieved by reduced interaction between the metal and the semiconductor, it can adversely affect the tunneling efficiency of the carriers through the interface, degrading the contact resistance, for example in the case of MIS contacts \cite{Yu2016}. Besides, a perfectly de-pinned contact may introduce additional device-to-device variability due to spurious change in doping.
\\
On the other hand, a de-pinned contact brings in an unprecedented advantage where ambipolar carrier injection is required. Two key applications are light-emitting diode (LED) and complementary metal oxide semiconductor field effect transistor (CMOSFET). In LED, one must seamlessly inject both electron and hole to the active layer to achieve high efficiency. In CMOSFET, it is crucial to achieve a low-resistance contact for both n-type and p-type source/drain for the n-FET and p-FET, respectively, using the same metal. In this work, we demonstrate efficient pulsed electroluminescence in 1L-WS$_2$ by contrasting the carrier injection from graphene and Au contacts. We also show strong ambipolar carrier injection by 2H-TaSe$_2$ contact in WSe$_2$ and WS$_2$ channel with an efficiency better than Au, strikingly, for both electrons and holes.
\\
\\
\textbf{Sources of ambiguity in SBH extraction:}
The back-gated FET (Figure \ref{Figures:Fig1}a) is a popular structure for the extraction of SBH ($\phi_B^0$) in layered semiconductors \cite{Das2013b,Gupta2019,Kaushik2014,Das2014,Liu2018,Kwon2017,Song2020,Chen2013,Dankert2014,Mahajan2019,Abraham2017,Zheng2019,Gourmelon2000}. The effective barrier height ($\phi_B$) is computed at different values of gate voltage (V$_g$) using Arrhenius equation. From the $\phi_B$ versus V$_g$  plot, the knee-point is identified as the flat-band condition where the transport mechanism switches from purely thermionic to tunneling or thermally assisted tunneling process and thus corresponds to the SBH of that interface. Such a qualitative estimation of the flat-band condition and the nature of carrier transport in the lateral device structure lead to several ambiguities: (1) In a typical field effect device, the band edge position linearly depends on V$_g$ in the sub-threshold regime. On the other hand, above the threshold voltage, the relation becomes sub-linear due to screening. Figure \ref{Figures:Fig1}(b) shows the simulated values of the position of the conduction band edge (with respect to the Fermi level) for an ultra-thin channel as a function of V$_g$ for varying channel doping. The results are obtained from coupled 1D Poisson-Schrodinger equations and, hence, independent of the explicit carrier injection mechanism from the source. This suggests that the knee-point does not necessarily reflect the flat-band condition for the source injection, and can be confounded by the gate electrostatics. For example, at low SBH, the knee point can occur when flat-band is not reached (panel A in Figure \ref{Figures:Fig1}c). On the other hand, for large SBH, the tunneling component can be low in the total current, and the extracted $\phi_B$ can still be well in the exponential region, even beyond the flat-band condition (panel B in Figure \ref{Figures:Fig1}c). This will push the knee-point beyond the true flat-band condition. (2) The source and drain contacts in the lateral architecture, as shown in Figure \ref{Figures:Fig1}a, act as a back-to-back diode \cite{Somvanshi2017a}. Further, the mixed-dimensional electron transport \cite{Kang2014a} in the source contact itself must be carefully addressed since this is not directly accounted for in the standard Arrhenius equation \cite{Somvanshi2017a}. (3) The change in the channel resistance with temperature due to a modulation in the carrier mobility further results in a significant ambiguity. This is particularly problematic for low SBH, where the channel resistance forms a large fraction of the total resistance. These ambiguities result in large variability in the extracted SBH, as depicted in Figure \ref{Figures:Fig1}d where we summarize the reported SBH values for various contacts to MoS$_2$.
\\
\\
\textbf{Proposed Method:}
To avoid these ambiguities, we propose a test structure where a multi-layer TMDC channel is vertically sandwiched between two asymmetric metal contacts (Figure \ref{Figures:Fig2}a), giving rise to a built-in electric field in the multi-layer (Figure \ref{Figures:Fig2}b). This technique provides an estimation of the SBH of both the contact interfaces simultaneously using three steps: (i) identification of the flat-band condition from the open-circuit voltage (V$_{oc}$) under photoexcitation; ii) extraction of SBH of one of the contacts ($\phi_{B1}^0$) from temperature ($T$) dependent current measurement at the flat-band condition using Arrhenius plot; iii) finally, extraction of the SBH of the other contact interface using $\phi_{B2}^0 = \phi_{B1}^0 +\mid V_{oc} \mid$. Note that, the Arrhenius equation remains directly applicable by enforcing a one-dimensional charge transport using the vertical structure instead of a lateral one. The channel length of the vertical device is only the thickness of the layered material. Thus, the total resistance is dominated by the source injection, eliminating any effect due to temperature-induced mobility variation.
\\
First, we explain how to identify the true flat-band condition. As the vertical device uses asymmetric contacts, light illumination on the device produces a finite short circuit current (I$_{sc}$) due to the presence of a built-in field in the active layer. The sign of $I_{sc}$ depends on the direction of the built-in field, which, in turn, is decided by the relative values of $\phi_{B1}^0$ and $\phi_{B2}^0$. Under light illumination, the net current is given by $I_{net} = I_{dark} + I_{ph}$ where $I_{dark}$ is the dark current and $I_{ph}$ is the photocurrent. The open-circuit voltage (V$_{oc}$) is identified as the voltage at which $I_{net}=0$, that is $I_{ph} = -I_{dark}$. It corresponds to the correct flat-band (zero field) condition only when $I_{ph}=0$, which in turn implies $I_{dark} = 0$ (Figure \ref{Figures:Fig2}c). We observe a significant reduction in the magnitude of V$_{oc}$ with an increase in $T$ (See Supporting Figure S1). This is attributed to a temperature-induced enhancement in $I_{dark}$ due to thermal injection over the barrier and thermally activated defects. For $I_{dark} \neq 0$, $I_{net}$ under illumination becomes zero at a finite opposite $I_{ph}$, thus V$_{oc}$ does not correspond to the true flat-band condition (Figure \ref{Figures:Fig2}d). To ensure a small $I_{dark}$, we thus measure the V$_{oc}$ at a low temperature ($T = 6.7$ K). In this context, it is also important to note that the effect of hot-carrier injection from metal contacts on determining correct flat-band condition can be neglected due to (i) inefficiency of the hot-carrier injection mechanism from metal contacts compared with the absorption by layered semiconductor, (ii) lack of drift of the hot carriers towards other electrode due to the absence of electric field under the flat-band condition, and (iii) partial cancellation of hot carrier injection effect because of the simultaneous injection of carriers from the top and bottom electrodes.
\\
We measure several samples for each of three different stacks employing multi-layer WS$_2$ (thickness $\sim$ 40-60 nm), namely, (i) Au-WS$_2$-FLG (D1) (ii) Au-WS$_2$-TaSe$_2$ (D2) and (iii) FLG-WS$_2$-TaSe$_2$ (D3) (see Methods in Supporting Information Note 1 for sample preparation and measurement details). The choice of the stacks allows us to estimate the SBH of each interface from two independent measurements. The temperature-dependent $I_{dark}$ characteristics (both forward and reverse sweeps) of representative samples (see Supporting Figure S2 for optical images and Raman characterization) from each stack are illustrated in Figure \ref{Figures:Fig2}e-g. In the same plots, we also show the current under light illumination at 6.7 K (red dashed trace) and the corresponding $V_{oc}$ values. In devices D1 and D2, the Au contact is biased, whereas in D3, the FLG contact is biased. The sign of V$_{oc}$ suggests that the photogenerated electrons in WS$_2$ are transferred to Au contact, and the corresponding equilibrium band diagram will have a bending as displayed in Figure \ref{Figures:Fig2}b.  Note that $I_{dark}$ at 6.7 K is negligible for all the devices at V$_{oc}$, leading to an accurate estimation of the flat-band condition.
\\
We now measure the device dark current at different $T$ keeping the bias same as in the low-temperature-flat-band condition, and the SBH is deduced from the slope of the Arrhenius plot, as shown in Figures \ref{Figures:Fig2}h-j. From Figures \ref{Figures:Fig2}h and \ref{Figures:Fig2}i, we estimate $\phi_B^0$(Au) to be 0.27 and 0.29 eV, respectively. Using Figures \ref{Figures:Fig2}e and \ref{Figures:Fig2}h, $\phi_B^0$(FLG) = $\phi_B^0$(Au) + $\mid V_{oc}\mid$ = (0.29 + 0.13) eV = 0.42 eV. Separately, $\phi_B^0$(FLG) is estimated to be 0.35 eV from Figure \ref{Figures:Fig2}j.  Similarly, $\phi_B^0$(TaSe$_2$) is estimated as 0.765 and 0.895 eV from Figures \ref{Figures:Fig2}f-i and \ref{Figures:Fig2}g-j, respectively.
\\
Figure \ref{Figures:Fig3} summarizes the extracted SBH values of Au, FLG, and TaSe$_2$ contacts to WS$_2$ and its relationship with work function (W$_m$) of the corresponding contact material. We use the work function values of FLG from the transfer characteristics we obtained in a top gated structure, which suggests the intrinsic nature of FLG in our experiment(See Supporting information Figure S3). Thus, we use a work function of 4.5 eV. For SnSe$_2$ and Au we extracted the work function values from Kelvin Probe Force Microscopy (KPFM) measurements \cite{Krishna2018}, whereas for TaSe$_2$, we use values from literature \cite{Zahin2017,Tsoutsou2016,Tsai2019}. The notation S in this figure which indicates the degree of the FLP, is expressed as:
\begin{equation}\label{eq:FLP}
S = \mid \frac{d(\phi_B^0)}{d(W_m)} \mid.
\end{equation}
$S=1$ corresponds to the ideal Schottky-Mott limit of de-pinning, while $S=0$ indicates a completely pinned Fermi level. Figure \ref{Figures:Fig3} shows that FLG exhibits a completely de-pinned contact ($S=1$), in good agreement with other recent works \cite{Yu2014,Chuang2014a,Leong2015,Farmanbar2016}. On the other hand, Au exhibits a high degree of pinning ($S=0.25$) close to the conduction band edge. The extracted average $S$ for TaSe$_2$ is $\sim$0.53 (with the highest obtained value being 0.72), which aligns close to the mid-gap of WS$_2$ and hence has the potential to be a superior ambipolar contact for CMOS applications, as discussed later. Results from a few other devices are summarized in Supporting Figure S4. The relatively large variability in the extracted SBH values could be intrinsically related to the vdW contact interfaces' de-pinned nature due to spurious changes in doping in WS$_2$, interface quality, and doping in the contact (for FLG).
\\
To further validate this method, we fabricated a separate TaSe$_2$-WS$_2$-SnSe$_2$ stack where SnSe$_2$ is a degenerately n-doped, highly conductive layered semiconductor \cite{Krishna2018}. The extracted SBH value of SnSe$_2$ is 0.71 eV (See Supporting Figure S4), which corresponds to $S = 0.8$, indicating a highly de-pinned contact.
\\
The proposed technique provides a direct way to estimate the SBH of two contact interfaces from a single measurement, fully adheres to the Arrhenius equation, avoids inaccuracy resulting from temperature-dependent mobility degradation, and provides an unambiguous estimation of the flat-band condition. As this method is governed by the work function difference between the two contacts, it is applicable for all the semiconducting layered materials irrespective of the nature of the band gap (direct or indirect) and absorption coefficient. Note that the V$_{oc}$ being extracted at low temperature can lead to a minor inaccuracy ($<40$ meV) in the extracted value of the SBH at room temperature due to a reduction in the bandgap from low temperature to room temperature ($\Delta E_g \sim 80$ meV as supported by temperature-dependent photoluminescence spectra in Supporting Figure S5). Also, one must ensure that tunneling induced dark current is negligible at V$_{oc}$ for the accurate estimation of the flat-band condition. SBH, being an interface property, is not expected to change significantly between few-layer and bulk materials. So, our estimated SBH using thicker TMDs should be equally applicable for few layers as well. However, ultra-thin (for example, a monolayer) samples will require additional caution due to possible tunnelling leakage.
\\
\\
\textbf{Pulsed electroluminescence - ambipolar versus unipolar contact:}
To further corroborate the values of SBH extracted for Au and FLG with WS$_2$, we prepare 1L-WS$_2$/Au and 1L-WS$_2$/FLG devices to demonstrate pulsed electroluminescence (EL). The schematic of the fabricated EL devices (optical images in Supporting Figure S6) are shown in Figure \ref{Figures:Fig4}a (top panel: 1L-WS$_2$/Au, bottom panel: 1L-WS$_2$/FLG). This is a two-terminal device where the source contact is grounded, and an ac voltage pulse is applied to the gate electrode. During the measurement, a 2 MHz voltage pulse with a peak-to-peak (-V$_p$ to +V$_p$) voltage of 5 V (rise/fall time  4/4 ns) is applied symmetrically about a dc offset voltage ($V_0$). As electrons and holes are injected from the source contact to the WS$_2$ layer alternatively, EL is produced at the transition edges near the source contact \cite{Lien2018,Paur}. The role of $V_0$ in the EL intensity is schematically explained in Figure \ref{Figures:Fig4}b for the case of a metal-contact with Fermi level pinned close to the conduction band. For $V_0$ = 0 V (top panel), when V$_g$ is at -V$_p$, the WS$_2$ layer is p-doped. The switching of the applied V$_g$ from -V$_p$ to +V$_p$ induces large tunneling of electrons from the source contact to WS$_2$. During the transition edge, some of the remaining holes in the valance band recombine with the incoming electrons, giving rise to an EL signal. The opposite mechanism generates another EL pulse when V$_g$ switches from +V$_p$ to -V$_p$. The measured steady-state EL intensity is the sum of the EL output arising from all such rising and falling transition edges.
\\
As the initial position of metal Fermi level is near the conduction band edge of WS$_2$, the limited availability of hole density limits the overall EL output. With $V_0<0$, the available hole density for recombination increases, in turn increasing the EL intensity, and reaches a maximum (middle panel of Figure \ref{Figures:Fig4}b) at $V_0=V_{max}$.  However, with further negative $V_0$, the injected electron density is reduced, eventually limiting the EL intensity (bottom panel). We thus expect a non-monotonic EL output as a function of $V_0$, and $V_0=V_{max}$ can be used as a probe to validate the relative SBH of Au and FLG qualitatively. Figure \ref{Figures:Fig4}c-e shows the EL spectra at different $V_0$, with an exciton peak around 2.01 eV at 300 K. Figure \ref{Figures:Fig4}f depicts the dependency of the peak intensity with $V_0$ for Au and FLG contacts. While for the FLG contact, the peak intensity occurs at a small negative $V_0$ ($V_{max} \sim -0.4$ V), for Au contact, one needs to apply a much larger negative voltage ($V_{max} < -2$ V). This suggests that Au is more strongly pinned close to the conduction band edge than FLG, in agreement with the extracted SBH values.
\\
\\
\textbf{TaSe$_2$ as an ambipolar vdW contact for CMOS:}
The estimated SBH indicates that TaSe$_2$, a prospective candidate for realizing weakly pinned vdW contact, aligns close to the mid-gap of WS$_2$ as shown in Figure \ref{Figures:Fig3}. This finding implies the possibility of using TaSe$_2$ to achieve pronounced ambipolar carrier injection, which allows the design optimization of TMD transistors for CMOS technology\cite{ren2019}. In order to demonstrate this, we fabricate top-gated lateral few-layer WS$_2$ transistor with asymmetric contacts, namely TaSe$_2$ and Au, as depicted in Figure \ref{Figures:Fig5}a (see Supporting Figure S7 in for sample details). The hBN layer beneath the channel layer minimizes any detrimental substrate effect on the electrical characteristics \cite{wang2019}. Using the transfer characteristics at different biasing configurations, we infer the carrier injection efficiency of TaSe$_2$ as a source contact with Au as a control. For an apple-to-apple comparison, we compare the drain current ($I_{ds}$) in the MOSFET mode ($V_{gs}>0$ and $V_{ds}>0$ for n-FET; $V_{gs}<0$ and $V_{ds}<0$ for p-FET), keeping Au and TaSe$_2$ as the source as illustrated in Figure \ref{Figures:Fig5}b. Further, to avoid the potential barrier at the drain end on the source injection efficiency, we compare the carrier injection level of one contact with the other at a relatively higher $V_{ds}$.
\\
With Au (TaSe$_2$) as the source contact, the colored regions of panels I and II (III and IV) in Figure \ref{Figures:Fig5}c depict the electron and hole injection characteristics of Au (TaSe$_2$) respectively. The corresponding output characteristics are provided in Supporting Figure S7. TaSe$_2$ exhibits ambipolar characteristics, while Au predominantly shows n-type behavior. The hole current in the p-FET mode is $2$x$10^{3}$-fold higher when TaSe$_2$ is used as a source compared with Au for an overdrive voltage ($|V_{gs}-V_{min}|$) of $3$ V at $|V_{ds}|=1$ V (indicated by black squares). Also, TaSe$_2$ exhibits enhanced electron injection compared to Au with 5-fold higher current in the n-FET mode at an overdrive of $5$ V and $|V_{ds}|=1$ V.
\\
We also demonstrate the ambipolar injection nature of TaSe$_2$ using a few-layer WSe$_2$ as the channel (see Supporting Figure S8 for sample details and characterization). The transfer characteristics at different biasing configurations are shown in Figure \ref{Figures:Fig5}d. Here, we find that Au shows ambipolar injection into WSe$_2$ with a slightly higher hole current than the electron current. These observations agree with the previous results seen from transferred Au contacts \cite{kong2020,liu2018vdW}. Compared to Au contact, TaSe$_2$ injection exhibits a 5-fold higher electron current in the n-FET mode and similar hole current in the p-FET mode, for an overdrive voltage of $4$ V. We also illustrate ambipolar characteristics of TaSe$_2$ with a monolayer WSe$_2$ channel, where TaSe$_2$ exhibits improved electron and hole injection than Au (see Supporting Figure S9).
\\
The striking fact that TaSe$_2$ as a source can inject both types of carriers (electrons for n-FET and holes for p-FET) more efficiently than Au suggests a weakly pinned nature of TaSe$_2$ contact leading to a possible modulation of the SBH by the gate voltage\cite{liu2016vdW}. We note that TaSe$_2$ is susceptible to oxidation which exhibits a-Se peak at $\sim$ $255 cm^{-1}$ in the Raman scattering of partially or completely oxidized flakes\cite{cartamil2015,sun2017}. However, during our device fabrication, we minimize the exposure of TaSe$_2$ to ambience by quick successive transfer of channel layer which avoids surface oxidation at TMD/TaSe$_2$ interface as discussed in Supporting information Note 2. Hence the role of interface TaO$_x$ is negligible on the depinning of TaSe$_2$ contact.\cite{li2017}
\\
\\
In summary, we demonstrated a simple technique to accurately estimate the Schottky barrier height at a vdW contact interface. We applied the technique to demonstrate that several types of layered contact materials, including semiconducting SnSe$_2$, semi-metallic few-layer graphene, and metallic TaSe$_2$ exhibit a universal de-pinned nature compared to bulk metals like Au. We exploited the ambipolar carrier injection from such de-pinned vdW contact to demonstrate two key applications: efficient pulsed electroluminescence and improved carrier injection for both pFET and nFET. The proposed SBH extraction technique and the universality of the de-pinned nature of vdW contacts should have intriguing prospects towards future nano-electronic devices.
\section*{SUPPORTING INFORMATION}
Supporting information is available.
\section*{ACKNOWLEDGMENTS}
This work was supported in part by the support of a grant from Indian Space Research Organization (ISRO), a grant from MHRD under STARS, grants under Ramanujan Fellowship and Nano Mission from the Department of Science and Technology (DST), Government of India, and support from MHRD, MeitY and DST Nano Mission through NNetRA. K.W. and T.T. acknowledge support from the Elemental Strategy Initiative conducted by the MEXT, Japan, Grant Number JPMXP0112101001, JSPS KAKENHI Grant Numbers JP20H00354 and the CREST(JPMJCR15F3), JST.
\section*{NOTES}
The authors declare no competing financial or non-financial interest.
\section*{Data Availability}
The data is available on reasonable request from the corresponding author.

\bibliography{reference}

\providecommand{\latin}[1]{#1}
\makeatletter
\providecommand{\doi}
  {\begingroup\let\do\@makeother\dospecials
  \catcode`\{=1 \catcode`\}=2 \doi@aux}
\providecommand{\doi@aux}[1]{\endgroup\texttt{#1}}
\makeatother
\providecommand*\mcitethebibliography{\thebibliography}
\csname @ifundefined\endcsname{endmcitethebibliography}
  {\let\endmcitethebibliography\endthebibliography}{}
\begin{mcitethebibliography}{57}
\providecommand*\natexlab[1]{#1}
\providecommand*\mciteSetBstSublistMode[1]{}
\providecommand*\mciteSetBstMaxWidthForm[2]{}
\providecommand*\mciteBstWouldAddEndPuncttrue
  {\def\EndOfBibitem{\unskip.}}
\providecommand*\mciteBstWouldAddEndPunctfalse
  {\let\EndOfBibitem\relax}
\providecommand*\mciteSetBstMidEndSepPunct[3]{}
\providecommand*\mciteSetBstSublistLabelBeginEnd[3]{}
\providecommand*\EndOfBibitem{}
\mciteSetBstSublistMode{f}
\mciteSetBstMaxWidthForm{subitem}{(\alph{mcitesubitemcount})}
\mciteSetBstSublistLabelBeginEnd
  {\mcitemaxwidthsubitemform\space}
  {\relax}
  {\relax}

\bibitem[Ng and Lynch(1987)Ng, and Lynch]{Ng1987}
Ng,~K.~K.; Lynch,~W.~T. {The Impact of Intrinsic Series Resistance on MOSFET
  Scaling}. \emph{IEEE Transactions on Electron Devices} \textbf{1987},
  \emph{34}, 503--511\relax
\mciteBstWouldAddEndPuncttrue
\mciteSetBstMidEndSepPunct{\mcitedefaultmidpunct}
{\mcitedefaultendpunct}{\mcitedefaultseppunct}\relax
\EndOfBibitem
\bibitem[Giubileo and {Di Bartolomeo}(2017)Giubileo, and {Di
  Bartolomeo}]{Giubileo2017}
Giubileo,~F.; {Di Bartolomeo},~A. {The role of contact resistance in graphene
  field-effect devices}. \emph{Progress in Surface Science} \textbf{2017},
  \emph{92}, 143--175\relax
\mciteBstWouldAddEndPuncttrue
\mciteSetBstMidEndSepPunct{\mcitedefaultmidpunct}
{\mcitedefaultendpunct}{\mcitedefaultseppunct}\relax
\EndOfBibitem
\bibitem[Taur and Ning(2013)Taur, and Ning]{Taur2007}
Taur,~Y.; Ning,~T.~H. \emph{{Fundamentals of Modern VLSI Devices}}; Cambridge
  University Press, 2013\relax
\mciteBstWouldAddEndPuncttrue
\mciteSetBstMidEndSepPunct{\mcitedefaultmidpunct}
{\mcitedefaultendpunct}{\mcitedefaultseppunct}\relax
\EndOfBibitem
\bibitem[Schulman \latin{et~al.}(2018)Schulman, Arnold, and Das]{Schulman2018}
Schulman,~D.~S.; Arnold,~A.~J.; Das,~S. {Contact engineering for 2D materials
  and devices}. \emph{Chemical Society Reviews} \textbf{2018}, \emph{47},
  3037--3058\relax
\mciteBstWouldAddEndPuncttrue
\mciteSetBstMidEndSepPunct{\mcitedefaultmidpunct}
{\mcitedefaultendpunct}{\mcitedefaultseppunct}\relax
\EndOfBibitem
\bibitem[Sze and Ng()Sze, and Ng]{Szeb}
Sze,~S.; Ng,~K.~K. \emph{{Physics of Semiconductor Devices}}; pp 141--142\relax
\mciteBstWouldAddEndPuncttrue
\mciteSetBstMidEndSepPunct{\mcitedefaultmidpunct}
{\mcitedefaultendpunct}{\mcitedefaultseppunct}\relax
\EndOfBibitem
\bibitem[Allain \latin{et~al.}()Allain, Kang, Banerjee, and Kis]{Allain2015}
Allain,~A.; Kang,~J.; Banerjee,~K.; Kis,~A. {Electrical contacts to
  two-dimensional semiconductors}. \emph{Nature Materials} 1195--1205\relax
\mciteBstWouldAddEndPuncttrue
\mciteSetBstMidEndSepPunct{\mcitedefaultmidpunct}
{\mcitedefaultendpunct}{\mcitedefaultseppunct}\relax
\EndOfBibitem
\bibitem[Kim \latin{et~al.}(2017)Kim, Moon, Lee, Choi, Ahmed, Nam, Cho, Shin,
  Park, and Yoo]{Kim2017}
Kim,~C.; Moon,~I.; Lee,~D.; Choi,~M.~S.; Ahmed,~F.; Nam,~S.; Cho,~Y.;
  Shin,~H.-j.; Park,~S.; Yoo,~W.~J. {Fermi Level Pinning at Electrical Metal
  Contacts of Monolayer Molybdenum Dichalcogenides}. \emph{ACS Nano}
  \textbf{2017}, \emph{11}, 1588--1596\relax
\mciteBstWouldAddEndPuncttrue
\mciteSetBstMidEndSepPunct{\mcitedefaultmidpunct}
{\mcitedefaultendpunct}{\mcitedefaultseppunct}\relax
\EndOfBibitem
\bibitem[Wang \latin{et~al.}(2020)Wang, Shao, Gong, and Shi]{Wang2020}
Wang,~Q.; Shao,~Y.; Gong,~P.; Shi,~X. {Metal-2D multilayered semiconductor
  junctions: layer-number dependent Fermi-level pinning}. \emph{Journal of
  Materials Chemistry C} \textbf{2020}, \emph{8}, 3113--3119\relax
\mciteBstWouldAddEndPuncttrue
\mciteSetBstMidEndSepPunct{\mcitedefaultmidpunct}
{\mcitedefaultendpunct}{\mcitedefaultseppunct}\relax
\EndOfBibitem
\bibitem[Gallacher \latin{et~al.}(2012)Gallacher, Velha, Paul, MacLaren,
  Myronov, and Leadley]{Gallacher2012}
Gallacher,~K.; Velha,~P.; Paul,~D.~J.; MacLaren,~I.; Myronov,~M.;
  Leadley,~D.~R. {Ohmic contacts to n-type germanium with low specific contact
  resistivity}. \emph{Applied Physics Letters} \textbf{2012}, \emph{100},
  022113\relax
\mciteBstWouldAddEndPuncttrue
\mciteSetBstMidEndSepPunct{\mcitedefaultmidpunct}
{\mcitedefaultendpunct}{\mcitedefaultseppunct}\relax
\EndOfBibitem
\bibitem[Nishimura \latin{et~al.}(2007)Nishimura, Kita, and
  Toriumi]{Nishimura2007}
Nishimura,~T.; Kita,~K.; Toriumi,~A. {Evidence for strong Fermi-level pinning
  due to metal-induced gap states at metal/germanium interface}. \emph{Applied
  Physics Letters} \textbf{2007}, \emph{91}, 123123\relax
\mciteBstWouldAddEndPuncttrue
\mciteSetBstMidEndSepPunct{\mcitedefaultmidpunct}
{\mcitedefaultendpunct}{\mcitedefaultseppunct}\relax
\EndOfBibitem
\bibitem[Lin \latin{et~al.}(2013)Lin, Yu, and Mohney]{Lin2013}
Lin,~J.~C.; Yu,~S.~Y.; Mohney,~S.~E. {Characterization of low-resistance ohmic
  contacts to n- and p-type InGaAs}. \emph{Journal of Applied Physics}
  \textbf{2013}, \emph{114}, 044504\relax
\mciteBstWouldAddEndPuncttrue
\mciteSetBstMidEndSepPunct{\mcitedefaultmidpunct}
{\mcitedefaultendpunct}{\mcitedefaultseppunct}\relax
\EndOfBibitem
\bibitem[Kim and Kim(1997)Kim, and Kim]{Yu1997}
Kim,~H.; Kim,~I. {PtTiPtAu and PdTiPtAu ohmic contacts to p-InGaAs}. Compound
  semiconductor 1997. Proceedings of the IEEE Twenty-Fourth International
  Symposium on Compound Semiconductors, San Diego, CA, USA. 1997; pp
  175--178\relax
\mciteBstWouldAddEndPuncttrue
\mciteSetBstMidEndSepPunct{\mcitedefaultmidpunct}
{\mcitedefaultendpunct}{\mcitedefaultseppunct}\relax
\EndOfBibitem
\bibitem[Zhang \latin{et~al.}(2019)Zhang, Le, Richter, and Hacker]{Zhang2019}
Zhang,~S.; Le,~S.~T.; Richter,~C.~A.; Hacker,~C.~A. {Improved contacts to
  p-type MoS$_2$ transistors by charge-transfer doping and contact
  engineering}. \emph{Applied Physics Letters} \textbf{2019}, \emph{115},
  073106\relax
\mciteBstWouldAddEndPuncttrue
\mciteSetBstMidEndSepPunct{\mcitedefaultmidpunct}
{\mcitedefaultendpunct}{\mcitedefaultseppunct}\relax
\EndOfBibitem
\bibitem[Zhang \latin{et~al.}(2013)Zhang, Ye, Yomogida, Takenobu, and
  Iwasa]{Zhang2013}
Zhang,~Y.~J.; Ye,~J.~T.; Yomogida,~Y.; Takenobu,~T.; Iwasa,~Y. {Formation of a
  stable p-n junction in a liquid-gated MoS$_2$ ambipolar transistor}.
  \emph{Nano Letters} \textbf{2013}, \emph{13}, 3023--3028\relax
\mciteBstWouldAddEndPuncttrue
\mciteSetBstMidEndSepPunct{\mcitedefaultmidpunct}
{\mcitedefaultendpunct}{\mcitedefaultseppunct}\relax
\EndOfBibitem
\bibitem[Chuang \latin{et~al.}(2014)Chuang, Battaglia, Azcatl, McDonnell, Kang,
  Yin, Tosun, Kapadia, Fang, Wallace, and Javey]{Chuang2014}
Chuang,~S.; Battaglia,~C.; Azcatl,~A.; McDonnell,~S.; Kang,~J.~S.; Yin,~X.;
  Tosun,~M.; Kapadia,~R.; Fang,~H.; Wallace,~R.~M.; Javey,~A. {MoS$_2$ P-type
  transistors and diodes enabled by high work function MoO$_x$ contacts}.
  \emph{Nano Letters} \textbf{2014}, \emph{14}, 1337--1342\relax
\mciteBstWouldAddEndPuncttrue
\mciteSetBstMidEndSepPunct{\mcitedefaultmidpunct}
{\mcitedefaultendpunct}{\mcitedefaultseppunct}\relax
\EndOfBibitem
\bibitem[Nipane \latin{et~al.}(2016)Nipane, Karmakar, Kaushik, Karande, and
  Lodha]{Nipane2016}
Nipane,~A.; Karmakar,~D.; Kaushik,~N.; Karande,~S.; Lodha,~S. {Few-Layer
  MoS$_2$ p-Type Devices Enabled by Selective Doping Using Low Energy
  Phosphorus Implantation}. \emph{ACS Nano} \textbf{2016}, \emph{10},
  2128--2137\relax
\mciteBstWouldAddEndPuncttrue
\mciteSetBstMidEndSepPunct{\mcitedefaultmidpunct}
{\mcitedefaultendpunct}{\mcitedefaultseppunct}\relax
\EndOfBibitem
\bibitem[Liu \latin{et~al.}(2016)Liu, Qu, Ryu, Ahmed, Yang, Lee, and
  Yoo]{Liu2016}
Liu,~X.; Qu,~D.; Ryu,~J.; Ahmed,~F.; Yang,~Z.; Lee,~D.; Yoo,~W.~J. {P-Type
  Polar Transition of Chemically Doped Multilayer MoS$_2$ Transistor}.
  \emph{Advanced Materials} \textbf{2016}, \emph{28}, 2345--2351\relax
\mciteBstWouldAddEndPuncttrue
\mciteSetBstMidEndSepPunct{\mcitedefaultmidpunct}
{\mcitedefaultendpunct}{\mcitedefaultseppunct}\relax
\EndOfBibitem
\bibitem[Liu \latin{et~al.}(2018)Liu, Guo, Zhu, Liao, Lee, Ding, Shakir,
  Gambin, Huang, and Duan]{Liu2018}
Liu,~Y.; Guo,~J.; Zhu,~E.; Liao,~L.; Lee,~S.~J.; Ding,~M.; Shakir,~I.;
  Gambin,~V.; Huang,~Y.; Duan,~X. {Approaching the Schottky-Mott limit in van
  der Waals metal-semiconductor junctions}. \emph{Nature} \textbf{2018},
  \emph{557}, 696--700\relax
\mciteBstWouldAddEndPuncttrue
\mciteSetBstMidEndSepPunct{\mcitedefaultmidpunct}
{\mcitedefaultendpunct}{\mcitedefaultseppunct}\relax
\EndOfBibitem
\bibitem[Song \latin{et~al.}(2020)Song, Sim, Kim, Kim, Oh, Na, Lee, Wang, Yan,
  Liu, Kwak, Chen, Cheong, Yoo, Lee, and Kwon]{Song2020}
Song,~S. \latin{et~al.}  {Wafer-scale production of patterned transition metal
  ditelluride layers for two-dimensional metal–semiconductor contacts at the
  Schottky–Mott limit}. \emph{Nature Electronics} \textbf{2020}, \emph{3},
  207--215\relax
\mciteBstWouldAddEndPuncttrue
\mciteSetBstMidEndSepPunct{\mcitedefaultmidpunct}
{\mcitedefaultendpunct}{\mcitedefaultseppunct}\relax
\EndOfBibitem
\bibitem[Mahajan \latin{et~al.}(2019)Mahajan, Kallatt, Dandu, Sharma, Gupta,
  and Majumdar]{Mehak2019b}
Mahajan,~M.; Kallatt,~S.; Dandu,~M.; Sharma,~N.; Gupta,~S.; Majumdar,~K. {Light
  emission from the layered metal 2H-TaSe$_2$ and its potential applications}.
  \emph{Communications Physics} \textbf{2019}, \emph{2}, 88\relax
\mciteBstWouldAddEndPuncttrue
\mciteSetBstMidEndSepPunct{\mcitedefaultmidpunct}
{\mcitedefaultendpunct}{\mcitedefaultseppunct}\relax
\EndOfBibitem
\bibitem[Krishna \latin{et~al.}(2018)Krishna, Kallatt, and
  Majumdar]{Krishna2018}
Krishna,~M.; Kallatt,~S.; Majumdar,~K. {Substrate effects in high gain , low
  operating voltage SnSe$_2$ photoconductor}. \emph{Nanotechnology}
  \textbf{2018}, \emph{29}, 035205\relax
\mciteBstWouldAddEndPuncttrue
\mciteSetBstMidEndSepPunct{\mcitedefaultmidpunct}
{\mcitedefaultendpunct}{\mcitedefaultseppunct}\relax
\EndOfBibitem
\bibitem[Singisetti \latin{et~al.}(2008)Singisetti, Wistey, Zimmerman,
  Thibeault, Rodwell, Gossard, and Bank]{Singisetti2008}
Singisetti,~U.; Wistey,~M.~A.; Zimmerman,~J.~D.; Thibeault,~B.~J.;
  Rodwell,~M.~J.; Gossard,~A.~C.; Bank,~S.~R. {Ultralow resistance in situ
  Ohmic contacts to InGaAs/InP}. \emph{Applied Physics Letters} \textbf{2008},
  \emph{93}, 183502\relax
\mciteBstWouldAddEndPuncttrue
\mciteSetBstMidEndSepPunct{\mcitedefaultmidpunct}
{\mcitedefaultendpunct}{\mcitedefaultseppunct}\relax
\EndOfBibitem
\bibitem[Shiraishi \latin{et~al.}(1994)Shiraishi, Furuhata, and
  Okamoto]{Shiraishi1994}
Shiraishi,~Y.; Furuhata,~N.; Okamoto,~A. {Infulence of
  metal/n-InAs/interlayer/n-GaAs structure on nonalloyed ohmic conytact
  resistance}. \emph{Journal of Applied Physics} \textbf{1994}, \emph{76},
  5099\relax
\mciteBstWouldAddEndPuncttrue
\mciteSetBstMidEndSepPunct{\mcitedefaultmidpunct}
{\mcitedefaultendpunct}{\mcitedefaultseppunct}\relax
\EndOfBibitem
\bibitem[Li \latin{et~al.}(2018)Li, Shin, Lee, Oh, and Lee]{Li2018}
Li,~M.; Shin,~G.; Lee,~J.; Oh,~J.; Lee,~H.~D. {Low contact resistance of NiGe/
  p -Ge by indium segregation during Ni germanidation}. \emph{AIP Advances}
  \textbf{2018}, \emph{8}, 065312\relax
\mciteBstWouldAddEndPuncttrue
\mciteSetBstMidEndSepPunct{\mcitedefaultmidpunct}
{\mcitedefaultendpunct}{\mcitedefaultseppunct}\relax
\EndOfBibitem
\bibitem[Spann \latin{et~al.}(2005)Spann, Anderson, Thornton, Harris, Thomas,
  and Tracy]{Spann2005}
Spann,~J.~Y.; Anderson,~R.~A.; Thornton,~T.~J.; Harris,~G.; Thomas,~S.~G.;
  Tracy,~C. {Characterization of nickel germanide thin films for use as
  contacts to p-channel germanium MOSFETs}. \emph{IEEE Electron Device Letters}
  \textbf{2005}, \emph{26}, 151--153\relax
\mciteBstWouldAddEndPuncttrue
\mciteSetBstMidEndSepPunct{\mcitedefaultmidpunct}
{\mcitedefaultendpunct}{\mcitedefaultseppunct}\relax
\EndOfBibitem
\bibitem[Das \latin{et~al.}(2013)Das, Chen, Penumatcha, and
  Appenzeller]{Das2013b}
Das,~S.; Chen,~H.~Y.; Penumatcha,~A.~V.; Appenzeller,~J. {High performance
  multilayer MoS$_2$ transistors with scandium contacts}. \emph{Nano Letters}
  \textbf{2013}, \emph{13}, 100--105\relax
\mciteBstWouldAddEndPuncttrue
\mciteSetBstMidEndSepPunct{\mcitedefaultmidpunct}
{\mcitedefaultendpunct}{\mcitedefaultseppunct}\relax
\EndOfBibitem
\bibitem[Chen \latin{et~al.}(2013)Chen, Odenthal, Swartz, Floyd, Wen, Luo, and
  Kawakami]{Chen2013}
Chen,~J.~R.; Odenthal,~P.~M.; Swartz,~A.~G.; Floyd,~G.~C.; Wen,~H.; Luo,~K.~Y.;
  Kawakami,~R.~K. {Control of Schottky barriers in single layer MoS$_2$
  transistors with ferromagnetic contacts}. \emph{Nano Letters} \textbf{2013},
  \emph{13}, 3106--3110\relax
\mciteBstWouldAddEndPuncttrue
\mciteSetBstMidEndSepPunct{\mcitedefaultmidpunct}
{\mcitedefaultendpunct}{\mcitedefaultseppunct}\relax
\EndOfBibitem
\bibitem[Yu \latin{et~al.}(2016)Yu, Schaekers, Demuynck, Barla, Mocuta,
  Horiguchi, Collaert, Thean, and {De Meyer}]{Yu2016}
Yu,~H.; Schaekers,~M.; Demuynck,~S.; Barla,~K.; Mocuta,~A.; Horiguchi,~N.;
  Collaert,~N.; Thean,~A. V.~Y.; {De Meyer},~K. {MIS or MS? Source/drain
  contact scheme evaluation for 7 nm Si CMOS technology and beyond}. 2016 16th
  International Workshop on Junction Technology, IWJT, Shanghai. 2016; pp
  19--24\relax
\mciteBstWouldAddEndPuncttrue
\mciteSetBstMidEndSepPunct{\mcitedefaultmidpunct}
{\mcitedefaultendpunct}{\mcitedefaultseppunct}\relax
\EndOfBibitem
\bibitem[Gupta \latin{et~al.}(2019)Gupta, Rortais, Ohshima, Ando, Endo, Miyata,
  and Shiraishi]{Gupta2019}
Gupta,~S.; Rortais,~F.; Ohshima,~R.; Ando,~Y.; Endo,~T.; Miyata,~Y.;
  Shiraishi,~M. {Monolayer MoS$_2$ field effect transistor with low Schottky
  barrier height with ferromagnetic metal contacts}. \emph{Scientific Reports}
  \textbf{2019}, \emph{9}, 1--7\relax
\mciteBstWouldAddEndPuncttrue
\mciteSetBstMidEndSepPunct{\mcitedefaultmidpunct}
{\mcitedefaultendpunct}{\mcitedefaultseppunct}\relax
\EndOfBibitem
\bibitem[Kaushik \latin{et~al.}(2014)Kaushik, Nipane, Basheer, Dubey, Grover,
  Deshmukh, and Lodha]{Kaushik2014}
Kaushik,~N.; Nipane,~A.; Basheer,~F.; Dubey,~S.; Grover,~S.; Deshmukh,~M.~M.;
  Lodha,~S. {Schottky barrier heights for Au and Pd contacts to MoS$_2$}.
  \emph{Applied Physics Letters} \textbf{2014}, \emph{105}, 113505\relax
\mciteBstWouldAddEndPuncttrue
\mciteSetBstMidEndSepPunct{\mcitedefaultmidpunct}
{\mcitedefaultendpunct}{\mcitedefaultseppunct}\relax
\EndOfBibitem
\bibitem[Das \latin{et~al.}(2014)Das, Prakash, Salazar, and
  Appenzeller]{Das2014}
Das,~S.; Prakash,~A.; Salazar,~R.; Appenzeller,~J. {Toward low-power
  electronics: Tunneling phenomena in transition metal dichalcogenides}.
  \emph{ACS Nano} \textbf{2014}, \emph{8}, 1681--1689\relax
\mciteBstWouldAddEndPuncttrue
\mciteSetBstMidEndSepPunct{\mcitedefaultmidpunct}
{\mcitedefaultendpunct}{\mcitedefaultseppunct}\relax
\EndOfBibitem
\bibitem[Kwon \latin{et~al.}(2017)Kwon, Lee, Yu, Lee, Cui, Hone, and
  Lee]{Kwon2017}
Kwon,~J.; Lee,~J.~y.; Yu,~Y.~j.; Lee,~C.~h.; Cui,~X.; Hone,~J.; Lee,~G.~h.
  {Thickness-dependent Schottky barrier height of MoS$_2$ field-effect
  transistors}. \emph{Nanoscale} \textbf{2017}, \emph{9}, 6151--6157\relax
\mciteBstWouldAddEndPuncttrue
\mciteSetBstMidEndSepPunct{\mcitedefaultmidpunct}
{\mcitedefaultendpunct}{\mcitedefaultseppunct}\relax
\EndOfBibitem
\bibitem[Dankert \latin{et~al.}(2014)Dankert, Langouche, Kamalakar, and
  Dash]{Dankert2014}
Dankert,~A.; Langouche,~L.; Kamalakar,~M.~V.; Dash,~S.~P. {High-performance
  molybdenum disulfide field-effect transistors with spin tunnel contacts}.
  \emph{ACS Nano} \textbf{2014}, \emph{8}, 476--482\relax
\mciteBstWouldAddEndPuncttrue
\mciteSetBstMidEndSepPunct{\mcitedefaultmidpunct}
{\mcitedefaultendpunct}{\mcitedefaultseppunct}\relax
\EndOfBibitem
\bibitem[Mahajan \latin{et~al.}(2019)Mahajan, Murali, Kawatra, and
  Majumdar]{Mahajan2019}
Mahajan,~M.; Murali,~K.; Kawatra,~N.; Majumdar,~K. {Gate-Controlled Large
  Resistance Switching Driven by Charge-Density Wave in 1T-TaS$_2$/2H-MoS$_2$
  Heterojunctions}. \emph{Physical Review Applied} \textbf{2019}, \emph{11},
  024031\relax
\mciteBstWouldAddEndPuncttrue
\mciteSetBstMidEndSepPunct{\mcitedefaultmidpunct}
{\mcitedefaultendpunct}{\mcitedefaultseppunct}\relax
\EndOfBibitem
\bibitem[Abraham and Mohney(2017)Abraham, and Mohney]{Abraham2017}
Abraham,~M.; Mohney,~S.~E. {Annealed Ag contacts to MoS$_2$ field-effect
  transistors}. \emph{Journal of Applied Physics} \textbf{2017}, \emph{122},
  115306\relax
\mciteBstWouldAddEndPuncttrue
\mciteSetBstMidEndSepPunct{\mcitedefaultmidpunct}
{\mcitedefaultendpunct}{\mcitedefaultseppunct}\relax
\EndOfBibitem
\bibitem[Zheng \latin{et~al.}(2019)Zheng, Lu, Liu, and Robertson]{Zheng2019}
Zheng,~S.; Lu,~H.; Liu,~D.; Robertson,~J. {Insertion of an ultrathin Al2O3
  interfacial layer for Schottky barrier height reduction in WS$_2$
  field-effect transistors}. \emph{Nanoscale} \textbf{2019}, \emph{11},
  4811--4821\relax
\mciteBstWouldAddEndPuncttrue
\mciteSetBstMidEndSepPunct{\mcitedefaultmidpunct}
{\mcitedefaultendpunct}{\mcitedefaultseppunct}\relax
\EndOfBibitem
\bibitem[Gourmelon \latin{et~al.}(2000)Gourmelon, Bern{\`{e}}de, Pouzet, and
  Marsillac]{Gourmelon2000}
Gourmelon,~E.; Bern{\`{e}}de,~J.~C.; Pouzet,~J.; Marsillac,~S. {Textured
  MoS$_2$ thin films obtained on tungsten: Electrical properties of the
  W/MoS$_2$ contact}. \emph{Journal of Applied Physics} \textbf{2000},
  \emph{87}, 1182--1186\relax
\mciteBstWouldAddEndPuncttrue
\mciteSetBstMidEndSepPunct{\mcitedefaultmidpunct}
{\mcitedefaultendpunct}{\mcitedefaultseppunct}\relax
\EndOfBibitem
\bibitem[Somvanshi \latin{et~al.}(2017)Somvanshi, Kallatt, Venkatesh, Nair,
  Gupta, Anthony, Karmakar, and Majumdar]{Somvanshi2017a}
Somvanshi,~D.; Kallatt,~S.; Venkatesh,~C.; Nair,~S.; Gupta,~G.; Anthony,~J.~K.;
  Karmakar,~D.; Majumdar,~K. {Nature of carrier injection in
  metal/2D-semiconductor interface and its implications for the limits of
  contact resistance}. \emph{Physical Review B} \textbf{2017}, \emph{96},
  205423\relax
\mciteBstWouldAddEndPuncttrue
\mciteSetBstMidEndSepPunct{\mcitedefaultmidpunct}
{\mcitedefaultendpunct}{\mcitedefaultseppunct}\relax
\EndOfBibitem
\bibitem[Kang \latin{et~al.}(2014)Kang, Liu, Sarkar, Jena, and
  Banerjee]{Kang2014a}
Kang,~J.; Liu,~W.; Sarkar,~D.; Jena,~D.; Banerjee,~K. {Computational study of
  metal contacts to monolayer transition-metal dichalcogenide semiconductors}.
  \emph{Physical Review X} \textbf{2014}, \emph{4}, 031005\relax
\mciteBstWouldAddEndPuncttrue
\mciteSetBstMidEndSepPunct{\mcitedefaultmidpunct}
{\mcitedefaultendpunct}{\mcitedefaultseppunct}\relax
\EndOfBibitem
\bibitem[Zahin(2017)]{Zahin2017}
Zahin,~A. {Schottky Barrier Heights at Two-Dimensional Metallic and
  Semiconducting Transition-Metal Dichalcogenide Interfaces}. Ph.D.\ thesis,
  2017\relax
\mciteBstWouldAddEndPuncttrue
\mciteSetBstMidEndSepPunct{\mcitedefaultmidpunct}
{\mcitedefaultendpunct}{\mcitedefaultseppunct}\relax
\EndOfBibitem
\bibitem[Tsoutsou \latin{et~al.}(2016)Tsoutsou, Aretouli, Tsipas,
  Marquez-Velasco, Xenogiannopoulou, Kelaidis, {Aminalragia Giamini}, and
  Dimoulas]{Tsoutsou2016}
Tsoutsou,~D.; Aretouli,~K.~E.; Tsipas,~P.; Marquez-Velasco,~J.;
  Xenogiannopoulou,~E.; Kelaidis,~N.; {Aminalragia Giamini},~S.; Dimoulas,~A.
  {Epitaxial 2D MoSe$_2$ (HfSe$_2$) Semiconductor/2D TaSe$_2$ Metal van der
  Waals Heterostructures}. \emph{ACS Applied Materials and Interfaces}
  \textbf{2016}, \emph{8}, 1836--1841\relax
\mciteBstWouldAddEndPuncttrue
\mciteSetBstMidEndSepPunct{\mcitedefaultmidpunct}
{\mcitedefaultendpunct}{\mcitedefaultseppunct}\relax
\EndOfBibitem
\bibitem[Tsai \latin{et~al.}(2019)Tsai, Liu, Liou, Chi, Tang, Wang, Ouyang,
  Chueh, Liu, Zhou, and Woon]{Tsai2019}
Tsai,~H.~S.; Liu,~F.~W.; Liou,~J.~W.; Chi,~C.~C.; Tang,~S.~Y.; Wang,~C.;
  Ouyang,~H.; Chueh,~Y.~L.; Liu,~C.; Zhou,~S.; Woon,~W.~Y. {Direct Synthesis of
  Large-Scale Multilayer TaSe2 on SiO2/Si Using Ion Beam Technology}. \emph{ACS
  Omega} \textbf{2019}, \emph{4}, 17536--17541\relax
\mciteBstWouldAddEndPuncttrue
\mciteSetBstMidEndSepPunct{\mcitedefaultmidpunct}
{\mcitedefaultendpunct}{\mcitedefaultseppunct}\relax
\EndOfBibitem
\bibitem[Yu \latin{et~al.}(2014)Yu, Lee, Ling, Santos, Shin, Lin, Dubey,
  Kaxiras, Kong, Wang, and Palacios]{Yu2014}
Yu,~L.; Lee,~Y.~H.; Ling,~X.; Santos,~E.~J.; Shin,~Y.~C.; Lin,~Y.; Dubey,~M.;
  Kaxiras,~E.; Kong,~J.; Wang,~H.; Palacios,~T. {Graphene/MoS$_2$ Hybrid
  technology for large-scale two-dimensional electronics}. \emph{Nano Letters}
  \textbf{2014}, \emph{14}, 3055--3063\relax
\mciteBstWouldAddEndPuncttrue
\mciteSetBstMidEndSepPunct{\mcitedefaultmidpunct}
{\mcitedefaultendpunct}{\mcitedefaultseppunct}\relax
\EndOfBibitem
\bibitem[Chuang \latin{et~al.}(2014)Chuang, Tan, Ghimire, Perera, Chamlagain,
  Cheng, Yan, Mandrus, Tom{\'{a}}nek, and Zhou]{Chuang2014a}
Chuang,~H.~J.; Tan,~X.; Ghimire,~N.~J.; Perera,~M.~M.; Chamlagain,~B.;
  Cheng,~M. M.~C.; Yan,~J.; Mandrus,~D.; Tom{\'{a}}nek,~D.; Zhou,~Z. {High
  mobility WSe$_2$ p - And n - Field-effect transistors contacted by highly
  doped graphene for low-resistance contacts}. \emph{Nano Letters}
  \textbf{2014}, \emph{14}, 3594--3601\relax
\mciteBstWouldAddEndPuncttrue
\mciteSetBstMidEndSepPunct{\mcitedefaultmidpunct}
{\mcitedefaultendpunct}{\mcitedefaultseppunct}\relax
\EndOfBibitem
\bibitem[Leong \latin{et~al.}(2015)Leong, Luo, Li, Khoo, Quek, and
  Thong]{Leong2015}
Leong,~W.~S.; Luo,~X.; Li,~Y.; Khoo,~K.~H.; Quek,~S.~Y.; Thong,~J.~T. {Low
  resistance metal contacts to MoS$_2$ devices with nickel-etched-graphene
  electrodes}. \emph{ACS Nano} \textbf{2015}, \emph{9}, 869--877\relax
\mciteBstWouldAddEndPuncttrue
\mciteSetBstMidEndSepPunct{\mcitedefaultmidpunct}
{\mcitedefaultendpunct}{\mcitedefaultseppunct}\relax
\EndOfBibitem
\bibitem[Farmanbar and Brocks(2016)Farmanbar, and Brocks]{Farmanbar2016}
Farmanbar,~M.; Brocks,~G. {Ohmic Contacts to 2D Semiconductors through van der
  Waals Bonding}. \emph{Advanced Electronic Materials} \textbf{2016}, \emph{2},
  1500405\relax
\mciteBstWouldAddEndPuncttrue
\mciteSetBstMidEndSepPunct{\mcitedefaultmidpunct}
{\mcitedefaultendpunct}{\mcitedefaultseppunct}\relax
\EndOfBibitem
\bibitem[Lien \latin{et~al.}(2018)Lien, Iii, and Han]{Lien2018}
Lien,~D.~h.; Iii,~J. W.~A.; Han,~K. {Large-area and bright pulsed
  electroluminescence in monolayer semiconductors}. \emph{Nature
  Communications} \textbf{2018}, \emph{9}, 1229\relax
\mciteBstWouldAddEndPuncttrue
\mciteSetBstMidEndSepPunct{\mcitedefaultmidpunct}
{\mcitedefaultendpunct}{\mcitedefaultseppunct}\relax
\EndOfBibitem
\bibitem[Paur \latin{et~al.}(2019)Paur, Molina-mendoza, and Mueller]{Paur}
Paur,~M.; Molina-mendoza,~A.~J.; Mueller,~T. {Electroluminescence from
  multi-particle exciton complexes in transition metal dichalcogenide
  semiconductors}. \emph{Nature Communications} \textbf{2019}, \emph{10},
  1709\relax
\mciteBstWouldAddEndPuncttrue
\mciteSetBstMidEndSepPunct{\mcitedefaultmidpunct}
{\mcitedefaultendpunct}{\mcitedefaultseppunct}\relax
\EndOfBibitem
\bibitem[Ren \latin{et~al.}(2019)Ren, Yang, Zhou, Mao, Han, and Zhou]{ren2019}
Ren,~Y.; Yang,~X.; Zhou,~L.; Mao,~J.-Y.; Han,~S.-T.; Zhou,~Y. Recent Advances
  in Ambipolar Transistors for Functional Applications. \emph{Advanced
  Functional Materials} \textbf{2019}, \emph{29}, 1902105\relax
\mciteBstWouldAddEndPuncttrue
\mciteSetBstMidEndSepPunct{\mcitedefaultmidpunct}
{\mcitedefaultendpunct}{\mcitedefaultseppunct}\relax
\EndOfBibitem
\bibitem[Wang \latin{et~al.}(2019)Wang, Tu, He, Wang, Yin, Cheng, Wang, Fang,
  and He]{wang2019}
Wang,~F.; Tu,~B.; He,~P.; Wang,~Z.; Yin,~L.; Cheng,~R.; Wang,~J.; Fang,~Q.;
  He,~J. Uncovering the Conduction Behavior of van der Waals Ambipolar
  Semiconductors. \emph{Advanced Materials} \textbf{2019}, \emph{31},
  1805317\relax
\mciteBstWouldAddEndPuncttrue
\mciteSetBstMidEndSepPunct{\mcitedefaultmidpunct}
{\mcitedefaultendpunct}{\mcitedefaultseppunct}\relax
\EndOfBibitem
\bibitem[Kong \latin{et~al.}(2020)Kong, Zhang, Tao, Zhang, Dang, Li, Feng,
  Liao, Duan, and Liu]{kong2020}
Kong,~L.; Zhang,~X.; Tao,~Q.; Zhang,~M.; Dang,~W.; Li,~Z.; Feng,~L.; Liao,~L.;
  Duan,~X.; Liu,~Y. Doping-free complementary WSe$_2$ circuit via van der Waals
  metal integration. \emph{Nature communications} \textbf{2020}, \emph{11},
  1--7\relax
\mciteBstWouldAddEndPuncttrue
\mciteSetBstMidEndSepPunct{\mcitedefaultmidpunct}
{\mcitedefaultendpunct}{\mcitedefaultseppunct}\relax
\EndOfBibitem
\bibitem[Liu \latin{et~al.}(2018)Liu, Guo, Zhu, Liao, Lee, Ding, Shakir,
  Gambin, Huang, and Duan]{liu2018vdW}
Liu,~Y.; Guo,~J.; Zhu,~E.; Liao,~L.; Lee,~S.-J.; Ding,~M.; Shakir,~I.;
  Gambin,~V.; Huang,~Y.; Duan,~X. Approaching the Schottky--Mott limit in van
  der Waals metal--semiconductor junctions. \emph{Nature} \textbf{2018},
  \emph{557}, 696--700\relax
\mciteBstWouldAddEndPuncttrue
\mciteSetBstMidEndSepPunct{\mcitedefaultmidpunct}
{\mcitedefaultendpunct}{\mcitedefaultseppunct}\relax
\EndOfBibitem
\bibitem[Liu \latin{et~al.}(2016)Liu, Stradins, and Wei]{liu2016vdW}
Liu,~Y.; Stradins,~P.; Wei,~S.-H. Van der Waals metal-semiconductor junction:
  Weak Fermi level pinning enables effective tuning of Schottky barrier.
  \emph{Science advances} \textbf{2016}, \emph{2}, e1600069\relax
\mciteBstWouldAddEndPuncttrue
\mciteSetBstMidEndSepPunct{\mcitedefaultmidpunct}
{\mcitedefaultendpunct}{\mcitedefaultseppunct}\relax
\EndOfBibitem
\bibitem[Cartamil-Bueno \latin{et~al.}(2015)Cartamil-Bueno, Steeneken,
  Tichelaar, Navarro-Moratalla, Venstra, van Leeuwen, Coronado, van~der Zant,
  Steele, and Castellanos-Gomez]{cartamil2015}
Cartamil-Bueno,~S.~J.; Steeneken,~P.~G.; Tichelaar,~F.~D.;
  Navarro-Moratalla,~E.; Venstra,~W.~J.; van Leeuwen,~R.; Coronado,~E.; van~der
  Zant,~H.~S.; Steele,~G.~A.; Castellanos-Gomez,~A. High-quality-factor
  tantalum oxide nanomechanical resonators by laser oxidation of TaSe 2.
  \emph{Nano Research} \textbf{2015}, \emph{8}, 2842--2849\relax
\mciteBstWouldAddEndPuncttrue
\mciteSetBstMidEndSepPunct{\mcitedefaultmidpunct}
{\mcitedefaultendpunct}{\mcitedefaultseppunct}\relax
\EndOfBibitem
\bibitem[Sun \latin{et~al.}(2017)Sun, Chen, Zhang, Sohrt, Zhao, Xu, Wang, Wang,
  Rossnagel, Gu, \latin{et~al.} others]{sun2017}
Sun,~L.; Chen,~C.; Zhang,~Q.; Sohrt,~C.; Zhao,~T.; Xu,~G.; Wang,~J.; Wang,~D.;
  Rossnagel,~K.; Gu,~L., \latin{et~al.}  Suppression of the Charge Density Wave
  State in Two-Dimensional 1T-TiSe2 by Atmospheric Oxidation. \emph{Angewandte
  Chemie International Edition} \textbf{2017}, \emph{56}, 8981--8985\relax
\mciteBstWouldAddEndPuncttrue
\mciteSetBstMidEndSepPunct{\mcitedefaultmidpunct}
{\mcitedefaultendpunct}{\mcitedefaultseppunct}\relax
\EndOfBibitem
\bibitem[Li \latin{et~al.}(2017)Li, Fan, Liu, Chen, Liu, Jia, Sun, Jiang, Han,
  Bouchiat, \latin{et~al.} others]{li2017}
Li,~X.-X.; Fan,~Z.-Q.; Liu,~P.-Z.; Chen,~M.-L.; Liu,~X.; Jia,~C.-K.;
  Sun,~D.-M.; Jiang,~X.-W.; Han,~Z.; Bouchiat,~V., \latin{et~al.}
  Gate-controlled reversible rectifying behaviour in tunnel contacted
  atomically-thin MoS 2 transistor. \emph{Nature communications} \textbf{2017},
  \emph{8}, 1--7\relax
\mciteBstWouldAddEndPuncttrue
\mciteSetBstMidEndSepPunct{\mcitedefaultmidpunct}
{\mcitedefaultendpunct}{\mcitedefaultseppunct}\relax
\EndOfBibitem
\end{mcitethebibliography}
\newpage
\newpage
\newpage
\newpage
\newpage
\AtEndDocument{

\section*{Supplementary material}

# Supporting Notes and Figures

## Note 1
## Methods

**Sample preparation for vertical devices:** The vertical devices used for the extraction of Schottky Barrier Height (SBH) are fabricated on a Si/SiO$_2$ substrate with pre-patterned Au contacts. Pre-deposited Au will act as one of the metal contact (M1) for devices D1 and D2. Multilayer WS$_2$ (40–60 nm) is identified on PDMS by optical contrast and then dry transferred to Au lines. Finally, the other contact, FLG/TaSe$_2$ for device D1/D2 is transferred precisely touching Au–WS$_2$ portion in one end and another Au line in other end. In the case of D3, FLG is transferred first on one of the Au lines, followed by the transfer of multilayer WS$_2$ in such a way that WS$_2$ touches only FLG and thus avoiding any contact with Au line. To conclude, TaSe$_2$ is transferred precisely on the Au-WS$_2$-FLG junction without touching either Au or FLG. Fabricating devices on pre-patterned substrate makes the whole process more efficient in terms of quality, time consumption and cost.

**Sample preparation for EL devices:** The devices for pulsed electro luminescent measurements are fabricated on a Si/SiO$_2$ substrate with pre-patterned Au contacts. For the fabrication of device with Au as source contact, monolayer WS$_2$ is mechanically exfoliated and identified using optical contrast on PDMS. Then, selected flake is dry transferred directly to one of the pre-patterned Au contacts. The thickness of the WS$_2$ is confirmed with PL measurement. Then WS$_2$ layer is capped with hBN and which also acts as gate dielectric. For gating the WS$_2$ channel, few layer graphene is used as gate electrode which is connected to another Au contact. For the device with FLG as source contact, we use back gated structure. So, at first, hBN is transferred to pre-patterned Au contact covering full width of the contact. Then monolayer WS$_2$ is transferred without touching Au contact. Finally, few layer graphene is transferred by ensuring a proper overlap with monolayer WS$_2$. Here, bottom Au electrode will act as gate contact and FLG acts as source contact. For both the devices,

gate pulse is applied between gate electrode and source contact and EL spectra is obtained from the $WS_2$ region which is close to the source contact.

**Sample preparation for Lateral FET devices:** The top-gated lateral FET devices are fabricated using pre-patterned Au contacts on $Si/SiO_2$ substrate. In order to avoid any effect of the substrate on the SBH of contacts and carrier injection characteristics, hBN ($\sim$ 10-15 nm) is transferred between source and drain pre-patterned Au contacts. Then, multilayer $TaSe_2$ ($\sim$ 20 nm) is transferred touching one of the Au contact which is followed by transer of few-layer TMD ($\sim$ 10 nm) to make contact with both Au and $TaSe_2$. For gating the TMD in both the channel and contact regions, hBN ($\sim$15-20 nm) and graphene layers are transferred successively on top of TMD covering both channel and contact regions with graphene touching another pre-patterned Au contact. Sample is heated after the transfer of every layer to ensure its better quality of contact. Materials are characterized by Raman spectroscopy at room temperature with 532 nm laser excitation.

**Electrical measurements:** All the electrical measurements are performed using Keithley 4200A-SCS Parameter Analyzer under vacuum level of $10^{-5}$ torr.

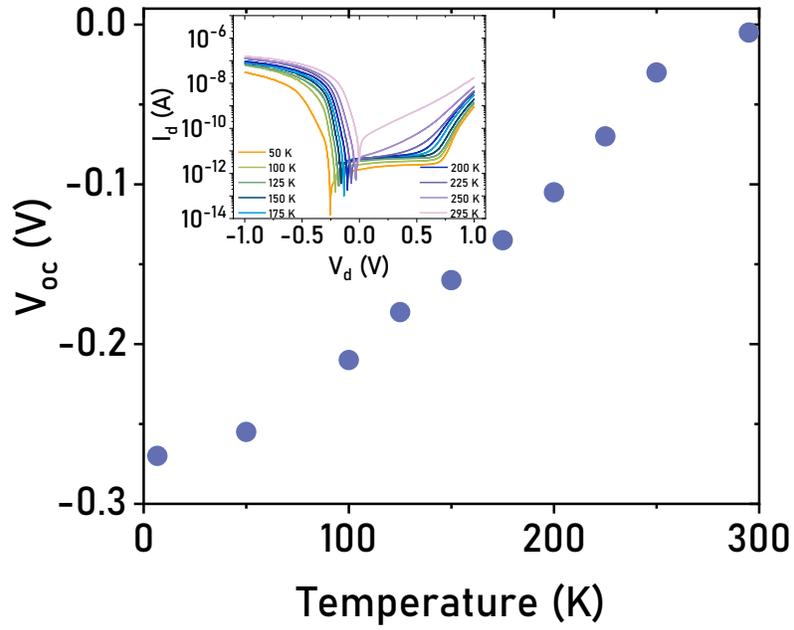

Figure S1: (a) Reduction in the magnitude of $V_{oc}$ observed for Au-WS$_2$-FLG stack with change in temperature resulting from the increase in the dark current.

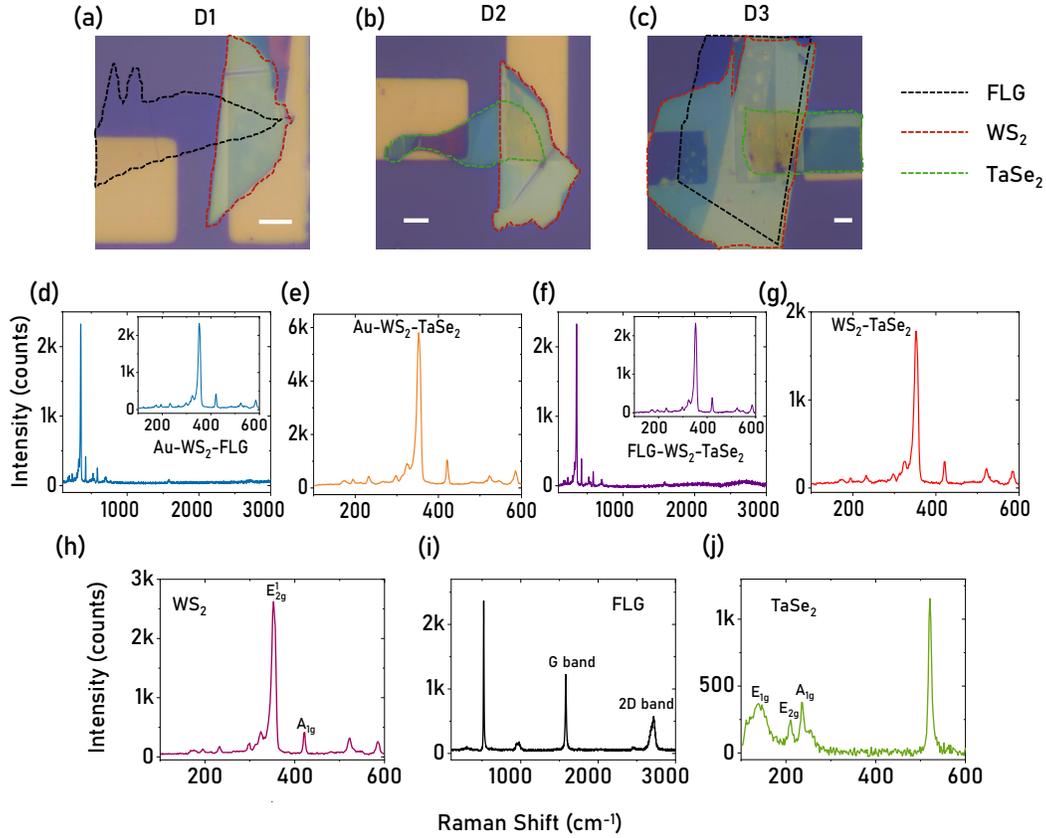

Figure S2: (a)-(c) Optical images of device (a) D1 (b) D2 and (c) D3. Scale bar is 5 $\mu$m. (d)-(j) Raman spectra at (d) Au-WS$_2$-FLG junction (e) Au-WS$_2$-TaSe$_2$ junction (f) FLG-WS$_2$-TaSe$_2$ junction (g) WS$_2$-TaSe$_2$ junction (h) isolated WS$_2$ portion (i)isolated FLG portion and (j) isolated TaSe$_2$ portion.

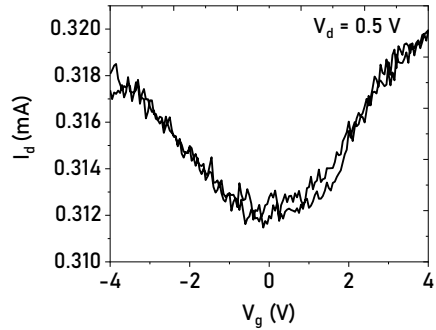

Figure S3: Transfer characteristics of FLG obtained in a top gated structure, where hBN is used as the top-gate dielectric. The minimum current occurs close to $V_g = 0$, suggesting negligible amount of doping in the FLG flake. We thus take the work function of FLG as 4.5 eV.

5

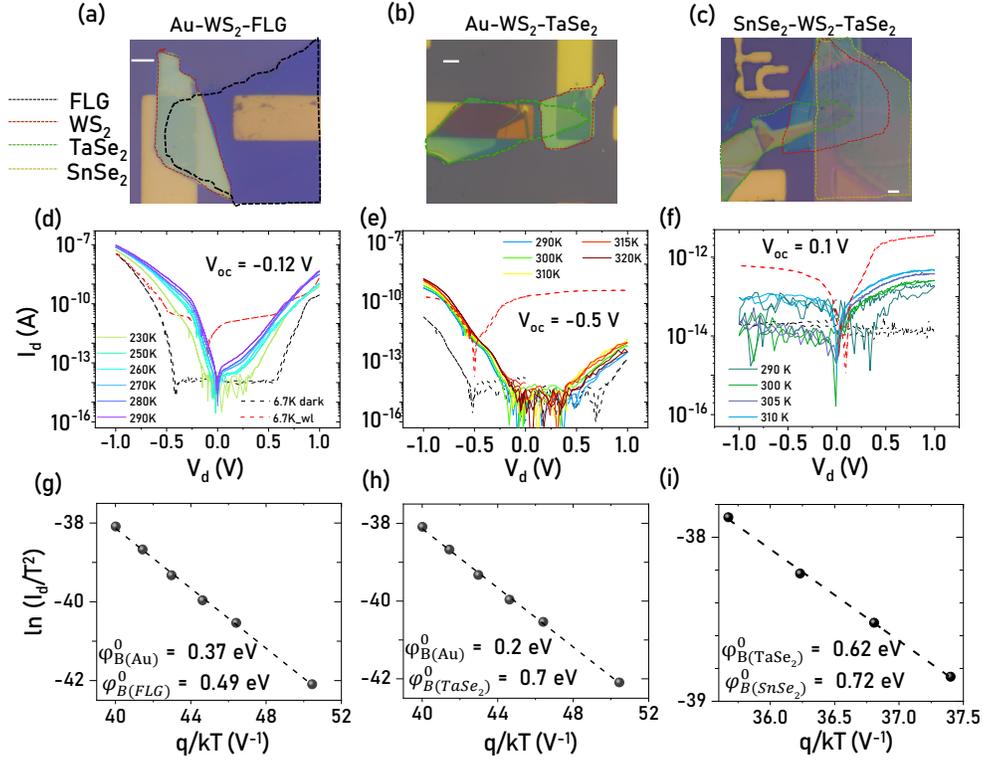

Figure S4: (a)-(c) Optical images of another set of vertical devices namely (a) Au-WS$_2$-FLG stack (DS1) (b) Au-WS$_2$-TaSe$_2$ stack (DS2) (c) SnSe$_2$-WS$_2$-TaSe$_2$ (DS3). Scale bar is 5 $\mu$m. (d)-(f) Current-Voltage characteristics under dark condition at different temperatures for the devices (d) DS1 (e) DS2 (f) DS3. The black and the red dashed traces indicate current with and without light, respectively, at $T = 6.7$ K. The $V_{oc}$ value at 6.7 K in each structure is indicated in the inset. (g)–(i) The corresponding Arrhenius plots for DS1-DS3 along with linear fits (dashed lines) to deduce the SBH of one of the contacts. The SBH of other contact is calculated by adding $|V_{oc}|$ to the first one.

6

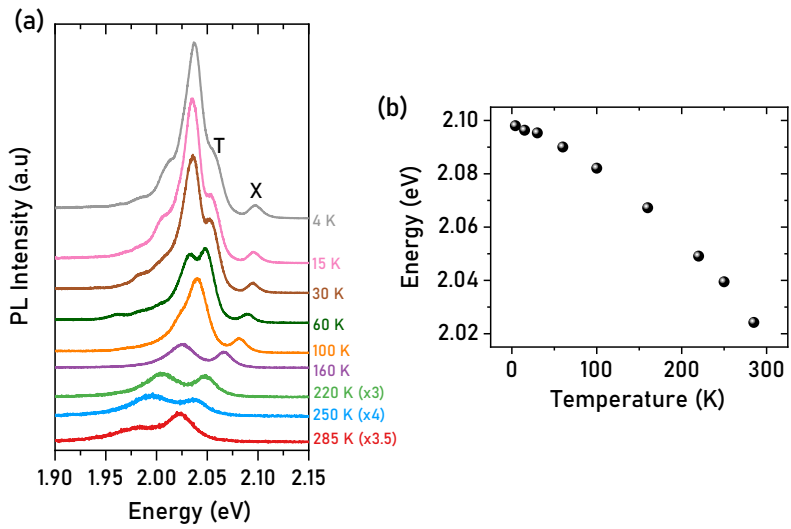

Figure S5: (a) Temperature dependent photoluminescence spectra of monolayer $WS_2$ where X and T are exciton and trion emission peaks respectively. (b) Red shift of exciton peak (X) with increase in the temperature.

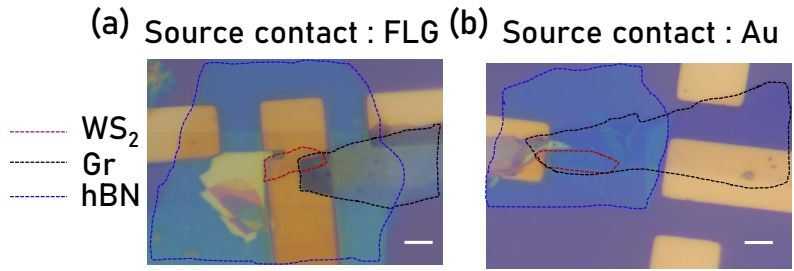

Figure S6: (a)-(b) Optical image of the EL devices where (a) FLG as source contact (b) Au as source contact. Scale bar is 5 $\mu$m.

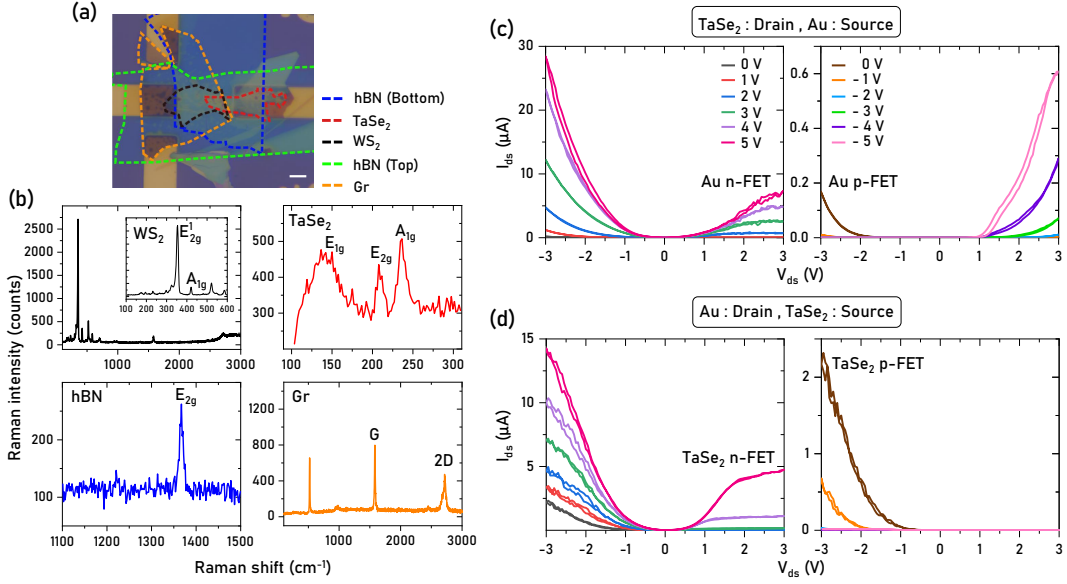

Figure S7: Top gated lateral few-layer WS$_2$ FET with Au and TaSe$_2$ contacts (a) Optical image of the device fabricated on pre-patterned Au contacts. Colored dotted lines highlight individual layers. Scale bar is 5 $\mu$m. (b) Raman characterization of WS$_2$ (in the channel region), TaSe$_2$, hBN and Graphene multilayers respectively. (c) Output characteristics of FET (in linear scale) with Au as source and TaSe$_2$ as drain under positive (left panel) and negative (right panel) gate bias. (d) Output characteristics of FET (in linear scale) with TaSe$_2$ as source and Au as drain under different gate bias.

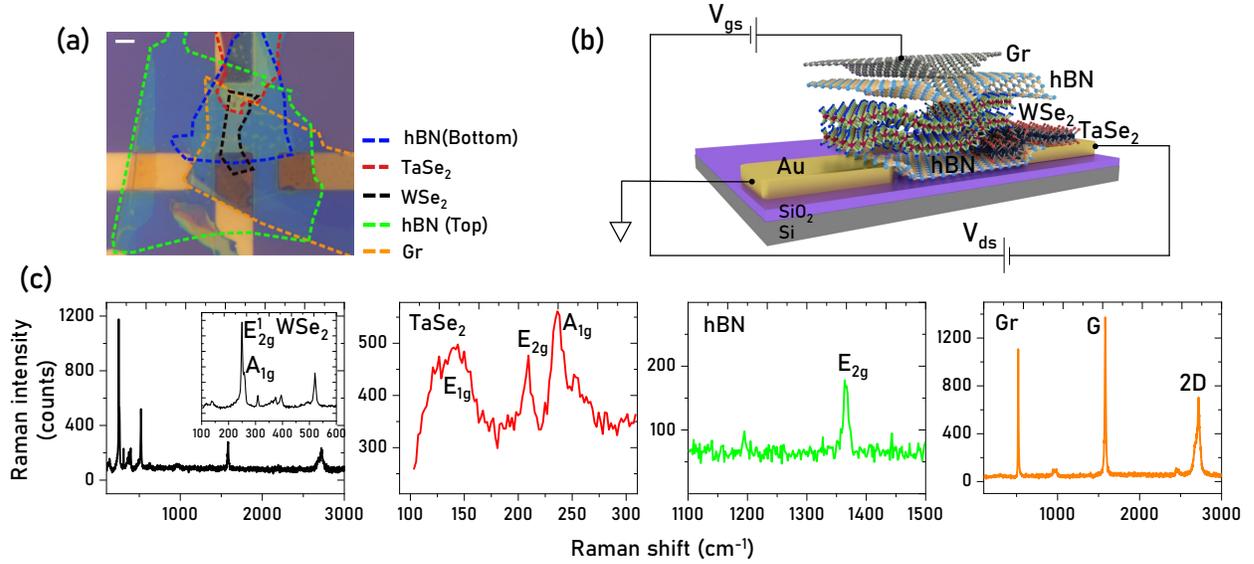

Figure S8: Top gated lateral FET with monolayer WSe$_2$ channel and asymmetric contacts (a) Optical image of the device highlighting regions of different layers along with monolayer and multilayer WSe$_2$. (b) Transfer characteristics of this FET (in log scale) at different biasing configurations. Colored regions in orange and green highlight the electron and hole injection characteristics of Au (I and II) and TaSe$_2$ (III and IV) respectively.

9

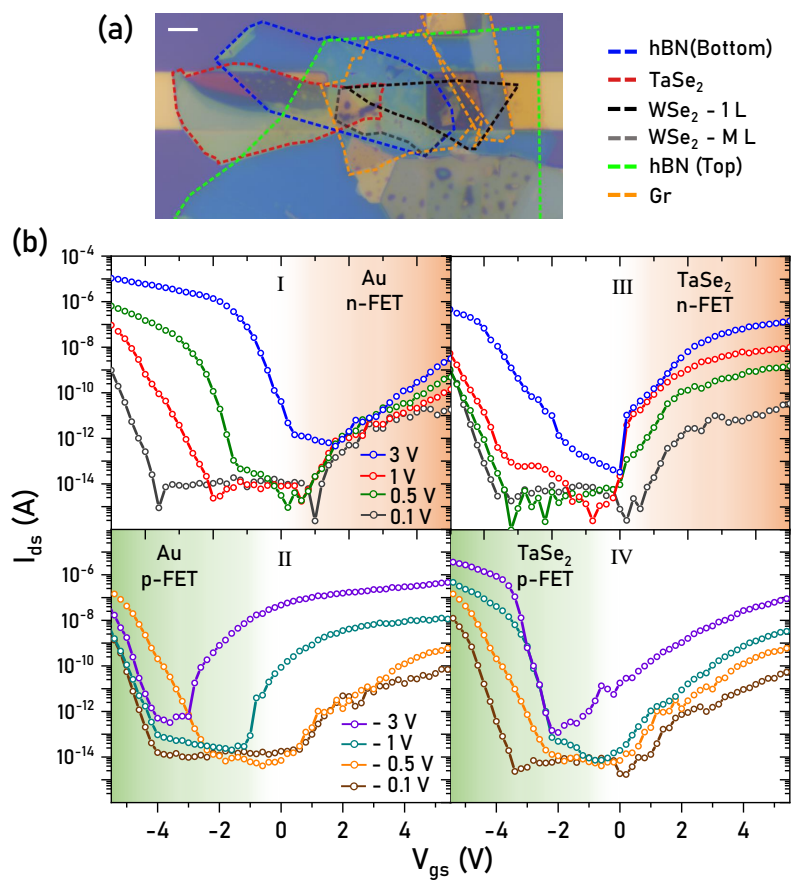

Figure S9: Top gated lateral few-layer WSe$_2$ FET with Au and TaSe$_2$ contacts (a) Optical image of the device showing the different layers. Scale bar is 5 $\mu$m. (b) Schematic of top-gated lateral FET with asymmetric contacts, Au and TaSe$_2$. The biasing configuration shown here is for TaSe$_2$ electron injection. (c) Raman characterization of WSe$_2$ (in the channel region), TaSe$_2$, hBN and graphene multilayers respectively.

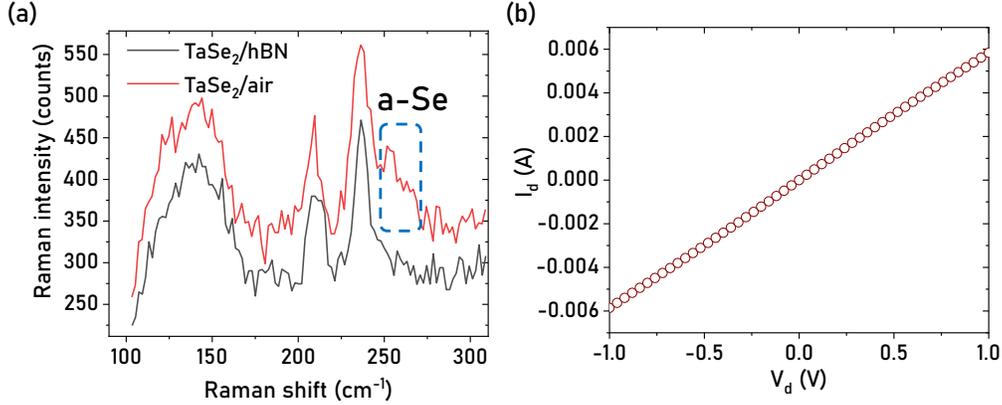

Figure S10: Stability of TaSe$_2$ (a) Raman characterization of air exposed and unexposed TaSe$_2$ portions which shows absence of oxidation in encapuslated TaSe$_2$. (b) Linear I-V characteristics of TaSe$_2$.

## Note 2

It has been been reported in literature that TaSe$_2$ is susceptible to oxidation at a relatively slower rate and a-Se peak at $\sim 255\ cm^{-1}$ is observed in the Raman scattering of partially or completely oxidized flakes.[1,2] However, in the FET devices which are fabricated on pre-patterned electrodes, we transfer the channel layer, WSe$_2$ or WS$_2$, within a few minutes after the exfoliation and transfer of TaSe$_2$. This minimizes the exposure of TaSe$_2$ to ambience and significantly avoids surface oxidation of TaSe$_2$ at the contact interface to WSe$_2$ or WS$_2$. Further, we encapsulate the FET devices with hBN layer at the top which isolates TaSe$_2$ from ambience.

We find a-Se peak in the Raman spectrum of TaSe$_2$ only when the TaSe$_2$ flake is exposed to air. However, such peak is found to be absent in the TaSe$_2$ portion encapsulated by hBN (see Figure S8a). These spectra were taken after three months of device fabrication and a very weak a-Se Raman peak at the air exposed TaSe$_2$ portion supports previous reports of a slow rate of oxidation of TaSe$_2$.[2] Moreover, any presence of surface oxide layer is usually reflected in I-V characteristics through a breakdown of the oxide layer. We find that TaSe$_2$ is highly conducting with linear I-V characteristics (see Figure S8b) which further indicates

the absence of any surface oxide layer. Hence the role of interface TaO$_x$ is negligible on the depinning of TaSe$_2$ contact.

# References

(1) Cartamil-Bueno, S. J.; Steeneken, P. G.; Tichelaar, F. D.; Navarro-Moratalla, E.; Venstra, W. J.; van Leeuwen, R.; Coronado, E.; van der Zant, H. S.; Steele, G. A.; Castellanos-Gomez, A. High-quality-factor tantalum oxide nanomechanical resonators by laser oxidation of TaSe 2. *Nano Research* **2015**, *8*, 2842–2849.

(2) Sun, L.; Chen, C.; Zhang, Q.; Sohrt, C.; Zhao, T.; Xu, G.; Wang, J.; Wang, D.; Rossnagel, K.; Gu, L., et al. Suppression of the Charge Density Wave State in Two-Dimensional 1T-TiSe2 by Atmospheric Oxidation. *Angewandte Chemie International Edition* **2017**, *56*, 8981–8985.

}
\end{document}